\documentclass[twocolumn]{aastex631}
\usepackage{amsmath}
\usepackage{booktabs}
\usepackage{float}
\usepackage{makecell}

\newcommand\comment[1]{\textcolor{black}{#1}}
\newcommand{\vgsr}{$V_{\mathrm{GSR}}$}
\newcommand{\dvgsr}{$\delta V_{\mathrm{GSR}}$}
\newcommand{\phione}{$\phi_1$}
\newcommand{\phitwo}{$\phi_2$}
\newcommand{\dphitwo}{$\delta\phi_2$}
\newcommand{\pmone}{$\mu_{\phi_1}$}
\newcommand{\pmtwo}{$\mu_{\phi_2}$}
\newcommand{\dpmone}{$\delta\mu_{\phi_1}$}
\newcommand{\dpmtwo}{$\delta\mu_{\phi_2}$}
\newcommand{\masyr}{mas~yr$^{-1}$}
\newcommand{\kms}{km~s$^{-1}$}
\newcommand{\feh}{[Fe/H]}

\begin{document}

\title{Characterizing the GD-1 Stream with DESI DR2 Data: Thin Stream and Hot Cocoon}

\shorttitle{GD-1 in DESI DR2}
\shortauthors{Jarvis, Li, Koposov et al.}

\correspondingauthor{Emma Jarvis, Ting S. Li}

\author[0009-0006-5612-7336]{Emma Jarvis}
\affiliation{David A. Dunlap Department of Astronomy \& Astrophysics, University of Toronto, 50 St. George Street, Toronto, ON, M5S 3H4}
\affiliation{Dunlap Institute for Astronomy \& Astrophysics, University of Toronto, 50 St George Street, Toronto, ON M5S 3H4, Canada}
\email{emma.jarvis@mail.utoronto.ca}

\author[0000-0002-9110-6163]{Ting S. Li}
\affiliation{David A. Dunlap Department of Astronomy \& Astrophysics, University of Toronto, 50 St. George Street, Toronto, ON, M5S 3H4}
\affiliation{Dunlap Institute for Astronomy \& Astrophysics, University of Toronto, 50 St George Street, Toronto, ON M5S 3H4, Canada}\email{ting.li@astro.utoronto.ca}

\author[0000-0003-2644-135X]{Sergey E. Koposov}
\affiliation{Institute for Astronomy, University of Edinburgh, Royal Observatory, Blackford Hill, Edinburgh EH9 3HJ, UK}
\affiliation{Institute of Astronomy, University of Cambridge, Madingley Road, Cambridge CB3 0HA, UK}

\author[0000-0002-7667-0081]{Raymond G. Carlberg}
\affiliation{David A. Dunlap Department of Astronomy \& Astrophysics, University of Toronto, 50 St. George Street, Toronto, ON, M5S 3H4}
\affiliation{Dunlap Institute for Astronomy \& Astrophysics, University of Toronto, 50 St George Street, Toronto, ON M5S 3H4, Canada}

\author[0000-0002-6257-2341]{Monica Valluri}
\affiliation{Department of Astronomy, University of Michigan, Ann Arbor, MI 48109, USA}

\author[0009-0008-1224-0382]{Nasser Mohammed}
\affiliation{David A. Dunlap Department of Astronomy \& Astrophysics, University of Toronto, 50 St. George Street, Toronto, ON, M5S 3H4}
\affiliation{Dunlap Institute for Astronomy \& Astrophysics, University of Toronto, 50 St George Street, Toronto, ON M5S 3H4, Canada}

\author{J.~Aguilar}
\affiliation{Lawrence Berkeley National Laboratory, 1 Cyclotron Road, Berkeley, CA 94720, USA}

\author[0000-0001-6098-7247]{S.~Ahlen}
\affiliation{Department of Physics, Boston University, 590 Commonwealth Avenue, Boston, MA 02215 USA}

\author[0000-0002-0084-572X]{Carlos Allende Prieto}
\affiliation{Instituto de Astrof\'{\i}sica de Canarias, C/ V\'{\i}a L\'{a}ctea, s/n, E-38205 La Laguna, Tenerife, Spain}
\affiliation{Departamento de Astrof\'{\i}sica, Universidad de La Laguna (ULL), E-38206, La Laguna, Tenerife, Spain}

\author[0000-0002-0740-1507]{Leandro {Beraldo e Silva}}
\affiliation{Observatório Nacional, Rio de Janeiro - RJ, 20921-400, Brasil}

\author[0000-0001-9712-0006]{D.~Bianchi}
\affiliation{Dipartimento di Fisica ``Aldo Pontremoli'', Universit\`a degli Studi di Milano, Via Celoria 16, I-20133 Milano, Italy}
\affiliation{INAF-Osservatorio Astronomico di Brera, Via Brera 28, 20122 Milano, Italy}

\author{D.~Brooks}
\affiliation{Department of Physics \& Astronomy, University College London, Gower Street, London, WC1E 6BT, UK}

\author[0000-0002-5689-8791]{Amanda Bystr\"om}
\affiliation{Institute for Astronomy, University of Edinburgh, Royal Observatory, Blackford Hill, Edinburgh EH9 3HJ, UK}

\author{T.~Claybaugh}
\affiliation{Lawrence Berkeley National Laboratory, 1 Cyclotron Road, Berkeley, CA 94720, USA}

\author[0000-0001-8274-158X]{A.~P.~Cooper}
\affiliation{Institute of Astronomy and Department of Physics, National Tsing Hua University, 101 Kuang-Fu Rd. Sec. 2, Hsinchu 30013, Taiwan}

\author[0000-0002-2169-0595]{A.~Cuceu}
\affiliation{Lawrence Berkeley National Laboratory, 1 Cyclotron Road, Berkeley, CA 94720, USA}

\author[0000-0002-1769-1640]{A.~de la Macorra}
\affiliation{Instituto de F\'{\i}sica, Universidad Nacional Aut\'{o}noma de M\'{e}xico,  Circuito de la Investigaci\'{o}n Cient\'{\i}fica, Ciudad Universitaria, Cd. de M\'{e}xico  C.~P.~04510,  M\'{e}xico}

\author[0000-0002-4928-4003]{Arjun~Dey}
\affiliation{NSF NOIRLab, 950 N. Cherry Ave., Tucson, AZ 85719, USA}

\author[0000-0002-5665-7912]{Biprateep~Dey}
\affiliation{Department of Astronomy \& Astrophysics, University of Toronto, Toronto, ON M5S 3H4, Canada}
\affiliation{Department of Physics \& Astronomy and Pittsburgh Particle Physics, Astrophysics, and Cosmology Center (PITT PACC), University of Pittsburgh, 3941 O'Hara Street, Pittsburgh, PA 15260, USA}

\author{P.~Doel}
\affiliation{Department of Physics \& Astronomy, University College London, Gower Street, London, WC1E 6BT, UK}

\author[0000-0002-2890-3725]{J.~E.~Forero-Romero}
\affiliation{Departamento de F\'isica, Universidad de los Andes, Cra. 1 No. 18A-10, Edificio Ip, CP 111711, Bogot\'a, Colombia}
\affiliation{Observatorio Astron\'omico, Universidad de los Andes, Cra. 1 No. 18A-10, Edificio H, CP 111711 Bogot\'a, Colombia}

\author[0000-0001-9632-0815]{E.~Gaztañaga}
\affiliation{Institut d'Estudis Espacials de Catalunya (IEEC), c/ Esteve Terradas 1, Edifici RDIT, Campus PMT-UPC, 08860 Castelldefels, Spain}
\affiliation{Institute of Cosmology and Gravitation, University of Portsmouth, Dennis Sciama Building, Portsmouth, PO1 3FX, UK}
\affiliation{Institute of Space Sciences, ICE-CSIC, Campus UAB, Carrer de Can Magrans s/n, 08913 Bellaterra, Barcelona, Spain}

\author[0000-0001-9852-9954]{Oleg Y. Gnedin}
\affiliation{University of Michigan, Ann Arbor, MI 48109, USA}

\author[0000-0003-3142-233X]{Satya~{Gontcho A Gontcho}}
\affiliation{University of Virginia, Department of Astronomy, Charlottesville, VA 22904, USA}

\author{G.~Gutierrez}
\affiliation{Fermi National Accelerator Laboratory, PO Box 500, Batavia, IL 60510, USA}

\author[0000-0002-6550-2023]{K.~Honscheid}
\affiliation{Center for Cosmology and AstroParticle Physics, The Ohio State University, 191 West Woodruff Avenue, Columbus, OH 43210, USA}
\affiliation{Department of Physics, The Ohio State University, 191 West Woodruff Avenue, Columbus, OH 43210, USA}
\affiliation{The Ohio State University, Columbus, 43210 OH, USA}

\author[0000-0003-0201-5241]{R.~Joyce}
\affiliation{NSF NOIRLab, 950 N. Cherry Ave., Tucson, AZ 85719, USA}

\author{R.~Kehoe}
\affiliation{Department of Physics, Southern Methodist University, 3215 Daniel Avenue, Dallas, TX 75275, USA}

\author[0000-0003-3510-7134]{T.~Kisner}
\affiliation{Lawrence Berkeley National Laboratory, 1 Cyclotron Road, Berkeley, CA 94720, USA}

\author[0000-0003-0853-8887]{Namitha Kizhuprakkat}
\affiliation{Institute of Astronomy and Department of Physics, National Tsing Hua University, 101 Kuang-Fu Rd. Sec. 2, Hsinchu 30013, Taiwan}

\author[0000-0001-6356-7424]{A.~Kremin}
\affiliation{Lawrence Berkeley National Laboratory, 1 Cyclotron Road, Berkeley, CA 94720, USA}

\author[0000-0002-2527-8899]{Mika Lambert}
\affiliation{Department of Astronomy \& Astrophysics, University of California, Santa Cruz, 1156 High Street, Santa Cruz, CA 95064, USA
}

\author[0000-0003-1838-8528]{M.~Landriau}
\affiliation{Lawrence Berkeley National Laboratory, 1 Cyclotron Road, Berkeley, CA 94720, USA}

\author[0000-0001-7178-8868]{L.~Le~Guillou}
\affiliation{Sorbonne Universit\'{e}, CNRS/IN2P3, Laboratoire de Physique Nucl\'{e}aire et de Hautes Energies (LPNHE), FR-75005 Paris, France}

\author[0000-0003-0105-9576]{Gustavo E. Medina}
\affiliation{David A. Dunlap Department of Astronomy \& Astrophysics, University of Toronto, 50 St. George Street, Toronto, ON, M5S 3H4}
\affiliation{Dunlap Institute for Astronomy \& Astrophysics, University of Toronto, 50 St George Street, Toronto, ON M5S 3H4, Canada}

\author[0000-0002-1125-7384]{A.~Meisner}
\affiliation{NSF NOIRLab, 950 N. Cherry Ave., Tucson, AZ 85719, USA}

\author{R.~Miquel}
\affiliation{Instituci\'{o} Catalana de Recerca i Estudis Avan\c{c}ats, Passeig de Llu\'{\i}s Companys, 23, 08010 Barcelona, Spain}
\affiliation{Institut de F\'{i}sica d’Altes Energies (IFAE), The Barcelona Institute of Science and Technology, Edifici Cn, Campus UAB, 08193, Bellaterra (Barcelona), Spain}

\author[0000-0001-9070-3102]{S.~Nadathur}
\affiliation{Institute of Cosmology and Gravitation, University of Portsmouth, Dennis Sciama Building, Portsmouth, PO1 3FX, UK}

\author[0000-0002-5758-150X]{Joan Najita}
\affiliation{NSF NOIRLab, 950 N Cherry Ave, Tucson, AZ, 85719
}

\author[0000-0003-3188-784X]{N.~Palanque-Delabrouille}
\affiliation{IRFU, CEA, Universit\'{e} Paris-Saclay, F-91191 Gif-sur-Yvette, France}
\affiliation{Lawrence Berkeley National Laboratory, 1 Cyclotron Road, Berkeley, CA 94720, USA}

\author[0000-0002-0644-5727]{W.~J.~Percival}
\affiliation{Department of Physics and Astronomy, University of Waterloo, 200 University Ave W, Waterloo, ON N2L 3G1, Canada}
\affiliation{Perimeter Institute for Theoretical Physics, 31 Caroline St. North, Waterloo, ON N2L 2Y5, Canada}
\affiliation{Waterloo Centre for Astrophysics, University of Waterloo, 200 University Ave W, Waterloo, ON N2L 3G1, Canada}

\author[0000-0001-7145-8674]{F.~Prada}
\affiliation{Instituto de Astrof\'{i}sica de Andaluc\'{i}a (CSIC), Glorieta de la Astronom\'{i}a, s/n, E-18008 Granada, Spain}

\author[0000-0001-6979-0125]{I.~P\'erez-R\`afols}
\affiliation{Departament de F\'isica, EEBE, Universitat Polit\`ecnica de Catalunya, c/Eduard Maristany 10, 08930 Barcelona, Spain}

\author[0000-0002-4900-2088]{Tian Qiu}
\affiliation{Department of Astronomy, School of Physics and Astronomy, and Key Laboratory for Particle Astrophysics and Cosmology (MOE)/Shanghai Key Laboratory for Particle Physics and Cosmology, Shanghai Jiao Tong University, Shanghai 200240, People's Republic of China}
\affiliation{State Key Laboratory of Dark Matter Physics, School of Physics and Astronomy, Shanghai Jiao Tong University, Shanghai 200240, China}

\author[0000-0001-5805-5766]{Alexander H.~Riley}
\affiliation{Lund Observatory, Division of Astrophysics, Department of Physics, Lund University, SE-221 00 Lund, Sweden
}

\author[0000-0002-6667-7028]{Constance M.~Rockosi}
\affiliation{Department of Astronomy \& Astrophysics, University of California, Santa Cruz, 1156 High Street, Santa Cruz, CA 95064, USA
}

\author{G.~Rossi}
\affiliation{Department of Physics and Astronomy, Sejong University, 209 Neungdong-ro, Gwangjin-gu, Seoul 05006, Republic of Korea}

\author[0000-0002-9646-8198]{E.~Sanchez}
\affiliation{CIEMAT, Avenida Complutense 40, E-28040 Madrid, Spain}

\author[0000-0002-7393-3595]{Nathan Sandford}
\affiliation{David A. Dunlap Department of Astronomy \& Astrophysics, University of Toronto, 50 St. George Street, Toronto, ON, M5S 3H4}

\author[0000-0002-3569-7421]{E.~F.~Schlafly}
\affiliation{Space Telescope Science Institute, 3700 San Martin Drive, Baltimore, MD 21218, USA}

\author{D.~Schlegel}
\affiliation{Lawrence Berkeley National Laboratory, 1 Cyclotron Road, Berkeley, CA 94720, USA}

\author[0000-0002-3461-0320]{J.~Silber}
\affiliation{Lawrence Berkeley National Laboratory, 1 Cyclotron Road, Berkeley, CA 94720, USA}

\author{D.~Sprayberry}
\affiliation{NSF NOIRLab, 950 N. Cherry Ave., Tucson, AZ 85719, USA}

\author[0000-0003-1704-0781]{G.~Tarl\'{e}}
\affiliation{University of Michigan, 500 S. State Street, Ann Arbor, MI 48109, USA}

\author{B.~A.~Weaver}
\affiliation{NSF NOIRLab, 950 N. Cherry Ave., Tucson, AZ 85719, USA}

\author[0000-0001-5381-4372]{R.~Zhou}
\affiliation{Lawrence Berkeley National Laboratory, 1 Cyclotron Road, Berkeley, CA 94720, USA}

\author[0000-0002-6684-3997]{H.~Zou}
\affiliation{National Astronomical Observatories, Chinese Academy of Sciences, A20 Datun Road, Chaoyang District, Beijing, 100101, P.~R.~China}

\collaboration{57}{(DESI Collaboration)}

\begin{abstract}

GD-1 is among the longest, coldest stellar streams in the Milky Way, making it an ideal target for probing dark matter substructure through dynamical heating. We present a catalog of \comment{608} spectroscopically confirmed GD-1 members from the first three years of Dark Energy Spectroscopic Instrument (DESI) observations. This constitutes the largest homogeneous spectroscopic sample of GD-1, doubling the number of members previously available only through heterogeneous compilations combining multiple surveys with different systematics. 
Using these data, we derive updated stream tracks in sky position, proper motion, and radial velocity that extend over $100^\circ$ of the stream. We apply a Gaussian mixture model to decompose the stream into a dynamically cold thin component ($\sigma_V = \comment{2.49\pm0.28}$~\kms, width $= \comment{0.23\pm0.01}^\circ$) and a kinematically hot cocoon ($\sigma_V = \comment{6.13\pm0.75}$~\kms, width $= \comment{2.18\pm0.17}^\circ$). 
The cocoon contains $\sim30\%$ of members and its velocity dispersion is consistent with $\sim11$~Gyr of heating by cold dark matter subhalos.
We also detect a large proper motion dispersion ($\comment{41.36\pm4.98}$~\kms) along the stream direction in the cocoon component. This feature indicates a significant line-of-sight distance spread in the cocoon, and its origin will be further explored in a forthcoming paper. These measurements demonstrate the power of DESI spectroscopy for characterizing the multi-component phase-space structure of stellar streams and constraining small-scale dark matter substructure.

\end{abstract}


\section{Introduction} \label{sec:intro}

The formation and evolution of galaxies are shaped by a continuous process of hierarchical assembly wherein smaller systems merge to form larger structures over cosmic time \citep{Press_1974, White_1978}. Galaxies like the Milky Way are expected to grow through the accretion and disruption of dwarf galaxies and globular clusters. This process is expected to leave behind tidal tails known as stellar streams \citep{Lynden-Bell-1995}. As a globular cluster is accreted onto a larger galaxy, the tidal field of the galaxy pulls stars away from the globular cluster into leading and trailing stellar streams \citep{Toomre_1972, Binney_2008}. The streams are thin and dynamically cold and will remain that way in a smooth potential with some increase in width due to orbital variations \citep{Erkal_Sanders_Belokurov_2016}. However, structure within the galactic potential, such as the visible disk, bar, and dwarf galaxies, as well as invisible dark-matter subhalos, perturb the orbits of stream stars to increase the velocity spread and induce density perturbations and gaps in the stream \citep{Ibata_2002, Siegal-Gaskins_2008, Carlberg_2013}. Therefore, streams are useful probes of dark matter \citep{Johnston_2002}.

There are several ways in which stellar streams can provide constraints on the properties of small-scale dark matter structures. For example, power spectra can be used to analyze density variations in streams by comparing the measured variations to those expected by baryonic structures in order to determine the contribution of dark matter subhalos to the density variations \citep{Banik_2021}. Similarly, dynamical modeling of a single subhalo perturber can be done to model the size and morphology of individual gaps along a stream \citep{Erkal_2016}. More recently, attention has shifted to the kinematic signatures of substructure: the cumulative effect of many subhalo encounters heats streams and drives their velocity distributions away from a single cold Gaussian, producing non-Gaussian wings and enhanced radial velocity dispersions \citep{Carlberg_2023, Carlberg2024}. As a result, measurements of the velocity distribution along observed streams, such as GD-1, can then be combined with perturbative models to infer the abundance and compactness of low-mass subhalos, providing a direct probe of the small-scale dark-matter distribution \citep{nibauer_2025, carlberg_2025}.

\begin{figure*}
    \centering
    \includegraphics[width=\linewidth]{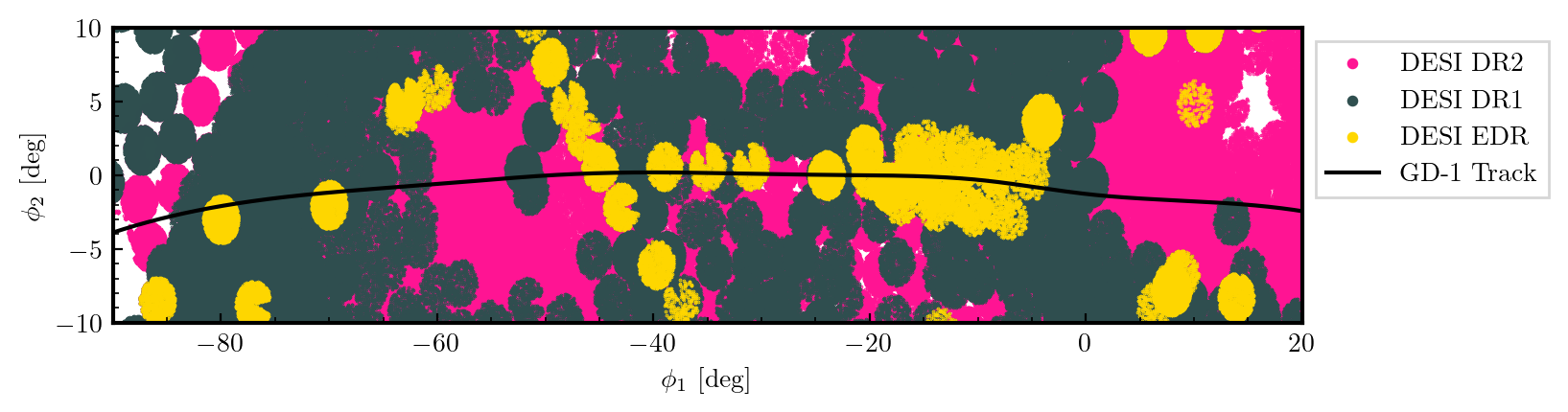}
    \caption{Comparison of DESI EDR (yellow; \citet{valluri_2025}), DR1 (grey), and DR2 (pink; this work, which includes all EDR and DR1 data) coverage in the GD-1 region. We note that the non-circular shape of the EDR tiles is due to one petal being temporarily out of operation. Overlaid in black is the GD-1 track derived in this work (see Section \ref{sec:phi2-sel}).}
    \label{fig:desi-releases}
\end{figure*}

Large scale surveys, such as {\it Gaia}, have led to the discovery of more than 100 stellar streams surrounding the Milky Way \citep{Mateu_2023, bonaca_2025}. Among these streams, GD-1 is one of the most prominent and well-studied. First discovered in 2006 using photometry from the Sloan Digital Sky Survey \citep{Grillmair_2006}, GD-1 is located in the Milky Way's halo and is believed to originate from a disrupted globular cluster due to its metallicity, width ($\sim20\arcmin$), and stellar mass (2~$\times 10^4$~M$_\odot$). The stream is very long, spanning at least $100^\circ$ at a heliocentric distance of $\sim$10~kpc, and has a retrograde orbit \citep{Koposov_2010}. GD-1 also has many noticeable density variations and deviations along its trajectory in the sky \citep{deBoer_2018}. For these reasons, it is an ideal target to study interactions with dark matter subhalos \citep{Banik_2021, carlberg_2025} or to constrain the Milky Way halo potential \citep{Koposov_2010, Malhan_Ibata_2019, Nibauer-bonaca2025}. 

\citet{carlberg_2025} shows that an evolving Cold Dark Matter (CDM) subhalo population acting for $\sim11$\,Gyr is able to heat GD-1 to radial velocity dispersions of 6.2\,\kms, matching the velocity dispersion measurement previously obtained using DESI EDR data \citep{valluri_2025}. While other theoretical mechanisms, such as interactions with giant molecular clouds, have been suggested to induce similar perturbations, simulations indicate that they do not match observed density variations in GD-1 \citep{Carlberg_2013}. Similarly, past encounters with known globular clusters have been ruled out as orbital integrations show that none have come close enough to GD-1 to produce its observed features \citep{Bonaca_2019, Doke_2022}. Resonant interactions with the Galactic bar can also induce perturbations, but such effects primarily influence prograde orbits \citep{Pearson_2017, Dillamore_2025}. For streams that orbit outside the disk, the effect of the disk and bar is weaker (particularly for a counter-rotating stream like GD-1), and encounters with dwarf galaxies are rare \citep{Bonaca_2019}.  Since GD-1 follows a retrograde orbit relative to the bar \citep{Price-Whelan_2018}, the interaction is expected to be minimal.

It has been well established that GD-1 contains many deviations from a thin stream track. In particular, \cite{Price-Whelan_2018} identify a ``spur" of stars extended away from the main stream track that is co-moving with the main stream. This spur has been shown to be consistent with a perturbation from a low-mass and compact dark matter subhalo \citep{Bonaca_2020}. A ``blob" of stars has also been identified below the main stream track \citep{Price-Whelan_2018, Bonaca_2019, tavangar_2025}. An alternative picture of these off-track components of GD-1 includes an extended distribution in \phione\ called a ``cocoon" \citep{Malhan_2019, valluri_2025}. This cocoon is a diffuse structure of stars surrounding the thin stream. \cite{valluri_2025} find that this structure has a width of $\sim460$\,pc, a velocity dispersion of $5-8$\,\kms, and that it extends over at least $30^\circ$ of the stream. \citet{ibata_2024} also find evidence of this cocoon-like envelope.

Studying these features of the stream requires a carefully constructed list of high-confidence member stars. Recent work has determined the membership probability and density of stars along the GD-1 stream. 811 GD-1 members were identified with the \texttt{STREAMFINDER} algorithm for {\it Gaia} Data Release 2 (DR2) \citep{Ibata_2020} and 1468 for {\it Gaia} Data Release 3 (DR3) \citep{ibata_2024}. Using data from {\it Gaia} and Pan-STARRS, \citet{Starkman_2025} identify 734 stars belonging to the main stream. More recently, \citet{tavangar_2025} identify 1689 high-probability GD-1 members. With data from the Dark Energy Spectroscopic Instrument (DESI) early data release observations \citep{DESI_2024}, \citet{valluri_2025} spectroscopically confirm 115 new member stars. \cite{tavangar_2025} also identify 353 high-probability members with radial velocities by combining data from several different spectroscopic surveys.

While current photometric studies have revealed off-track components in GD-1, the full picture of this stream remains incomplete. Additional spectroscopic observations are essential to identify members, precisely study the velocity distribution, and distinguish between the competing explanations for GD-1's off-stream features. With more kinematic data, we can more precisely calculate the velocity dispersion along the stream and place stronger constraints on the nature of dark matter subhalos and the role they play in perturbing stellar streams.

In this paper, we introduce the spectroscopically identified stars associated with the GD-1 stellar stream from the first three years of DESI observations. In Section \ref{sec:data} we describe the DESI data in the observed GD-1 region. In Section \ref{sec:members} we discuss our initial selection of GD-1 members based on cuts in the stellar color-magnitude diagram (CMD), in {\it Gaia} proper motions as well as in DESI radial velocities, and metallicities [Fe/H]. In Section \ref{sec:GMM} we study the thin and broad (``cocoon") components of the stream. In Section \ref{sec:discussion} we discuss our results. Finally, in Section \ref{sec:summary} we summarize our findings.

\section{DESI observations and DR2 Data} \label{sec:data}

The Dark Energy Spectroscopic Instrument (DESI) is a robotic, fibre-fed, highly-multiplexed spectrograph installed on the Mayall 4-m telescope at Kitt Peak National Observatory \citep{DESI_2022}. DESI uses 5,000 robotic fibre positioners over a $\sim 3^\circ$-diameter field of view to obtain optical spectra for millions of targets \citep{desi_2016, DESI_2016b, Miller_2024, Poppett_2024}. The DESI survey is an eight-year program \citep{Schlafly_2023} covering $\sim 17{,}000~\mathrm{deg}^2$ of sky and is designed to obtain spectra for $\sim 63$ million galaxies and quasars \citep{Guy_2023} and more than $10$ million Milky Way stars. Early results from DESI have already set exciting constraints on dark energy \citep{DESI_Collaboration_2024_VII, Collaboration_2025_II}.

DESI observations are organized into \emph{surveys} (e.g.\ commissioning/\texttt{CMX}, Survey Validation \texttt{SV1}--\texttt{SV3}, \texttt{main}, and \texttt{special}) and \emph{programs} (primarily \texttt{dark}, \texttt{bright}, and \texttt{backup}) that correspond to different weather conditions and science goals \citep{DESI_2024,Koposov_2024}. The main cosmological DESI survey is conducted in the dark program, while the bright program is used primarily for the Bright Galaxy Survey (BGS; \citealt{Hahn_2023}) and the Milky Way Survey (MWS; \citealt{Cooper_2023}). The backup program is a separate DESI survey that operates in poor conditions, targeting bright stars ($12 \lesssim G \lesssim 19$) over an extended footprint using {\it Gaia}-based selection \citep{Dey_2025}. All three programs are carried out during each survey phase, and spectra are coadded and catalogued separately for every combination of survey and program in the DESI data releases \citep{DESI_2025,Koposov_2025}.

Within the MWS, targets are selected into a set of well-defined classes encoded in the bitmasks \texttt{DESI\_TARGET}, \texttt{MWS\_TARGET}, and \texttt{SCND\_TARGET} \citep{Myers_2023,Cooper_2023,Koposov_2024}. For the main survey, the dominant MWS classes in bright time are the magnitude-limited main sample and several higher-priority categories. The main sample is subdivided primarily into \texttt{MWS\_MAIN\_BLUE}, \texttt{MWS\_MAIN\_RED}, and \texttt{MWS\_BROAD}, which together form a largely magnitude-limited selection in the range $16<r<19$ (in extinction-corrected DESI Legacy Survey $r$ band). Additional high-priority classes include white dwarfs, blue horizontal branch (BHB) stars, nearby stars within $\sim 100$ pc, and various secondary target categories observed in both bright and dark time \citep{Koposov_2024,Koposov_2025}. During Survey Validation, the MWS target definitions evolved and additional experimental target classes were introduced; a comprehensive description is given in \citet{AllendePrieto_2020, Cooper_2023, Koposov_2024}.

The DESI Early Data Release (EDR; \citealt{DESI_2024}) comprises commissioning and survey validation data obtained between December 2020 and June 2021. The EDR MWS Value-Added Catalogue (MWS VAC) \citep{Koposov_2024} provides a homogeneous set of radial velocities and stellar parameters for Milky Way targets, including the initial DESI GD-1 sample analyzed by \citet{valluri_2025}. The first full DESI data release (DR1; \citealt{DESI_2025}) extends this to 13 months of observations, including reprocessed SV data and early main-survey spectra. The DR1 stellar catalogue \citep{Koposov_2025} describes the combined stellar sample across all surveys and programs, including the growing contribution from the Milky Way Backup Program \citep{Dey_2025}. In this work we use the DR1 data together with additional Year~2 and Year~3 DESI observations (which will form part of DR2, to be released in Spring 2027), processed with the same pipelines and MWS VAC infrastructure.

Figure~\ref{fig:desi-releases} compares the DESI GD-1 coverage from the EDR (as analysed by \citealt{valluri_2025}), from DR1 (catalog publicly available with detailed description in \citealt{Koposov_2025}), and from DR2 (Years~1--3) used in this work. The progression from EDR to DR2 substantially increases the spatial extent, sampling density, and program diversity (\texttt{bright},  \texttt{dark}, and  \texttt{backup}) of DESI spectroscopy along GD-1, enabling the more detailed chemodynamical analysis presented in the following sections. In Appendix~\ref{sec:appendix_completeness} we provide further details on the DESI survey coverage of GD-1 across different programs and characterize the spectroscopic completeness of our sample. Figure \ref{fig:desi-releases} is in a coordinate system where $\phi_1$ is the angle along the stream and $\phi_2$ is the angle perpendicular to it. The transformation from equatorial coordinates $(\alpha, \delta)$  to $(\phi_1,\phi_2)$  follows the great-circle rotation defined in \citet{Koposov_2010}.
Most targets in the GD-1 region were originally selected from the DESI Legacy Surveys \citep{decals_2019}, from which we adopt the extinction-corrected $g$- and $r$-band photometry using the $E(B-V)$ values from \cite{Schlegel_1998}. These photometric catalogues were cross-matched with {\it Gaia}~DR3 \citep{Lindegren_2021a, gaia_dr3_2023} to obtain parallaxes and proper motions. Parallaxes were zero-point corrected following the prescriptions in \citet{Lindegren_2021a}.

We correct the {\it Gaia} DR3 proper motions for solar reflex motion using the \texttt{Galactocentric} frame from \texttt{astropy} \citep{Astropy_Collaboration_2022} with the \texttt{v4.0} parameter set, which adopts a Galactocentric distance of $R_\odot = 8.122$\,kpc, and a solar velocity of
$(U_\odot, V_\odot, W_\odot) = (12.9, 245.6, 7.78)$\,km\,s$^{-1}$.
The reflex correction is applied assuming an initial distance track for GD-1 from \citet{valluri_2025}, and then updated using the refined distance track derived in this work
(Section~\ref{sec:dist-track}).
The proper motions are rotated into the stream-aligned frame to obtain $(\mu_{\phi_1},\mu_{\phi_2})$, and the corresponding uncertainties and covariances are propagated through the transformation.

Radial velocities and stellar atmospheric parameters for all DESI spectra are measured with the \texttt{RVSpecFit} pipeline \citep{Koposov_2019}, which fits model spectra to the coadded DESI data in each survey/program combination; see more details in \citet{Koposov_2024, Koposov_2025}. Distances to individual stars are estimated using the \texttt{rvsdistnn} method (Koposov et al., in prep.), a neural network that maps the spectroscopic parameters from \texttt{RVSpecFit} with photometric information to distance estimates. We do not use these individual stellar distances directly in our dynamical modelling; instead, we employ them as an independent consistency check on our spectroscopic membership selection.

\section{Stream Membership Selection} \label{sec:members}

In the following section, we describe our criteria for selecting member stars of GD-1. This is done with kinematic, color-magnitude and metallicity cuts to the stars in the region of GD-1. To perform these cuts, we derive new empirical distance (Section \ref{sec:dist-track}) and CMD tracks (Section \ref{sec:cmd}). We then fit new proper motion (Section \ref{sec:pm_sel}) and \phitwo\ (Section \ref{sec:phi2-sel}) stream tracks using {\it Gaia} stars and derive a new \vgsr\ track (Section \ref{sec:rv-sel}) using the radial velocities of stars in DESI DR2.

\begin{figure*}
    \centering
    \includegraphics[width=\linewidth]{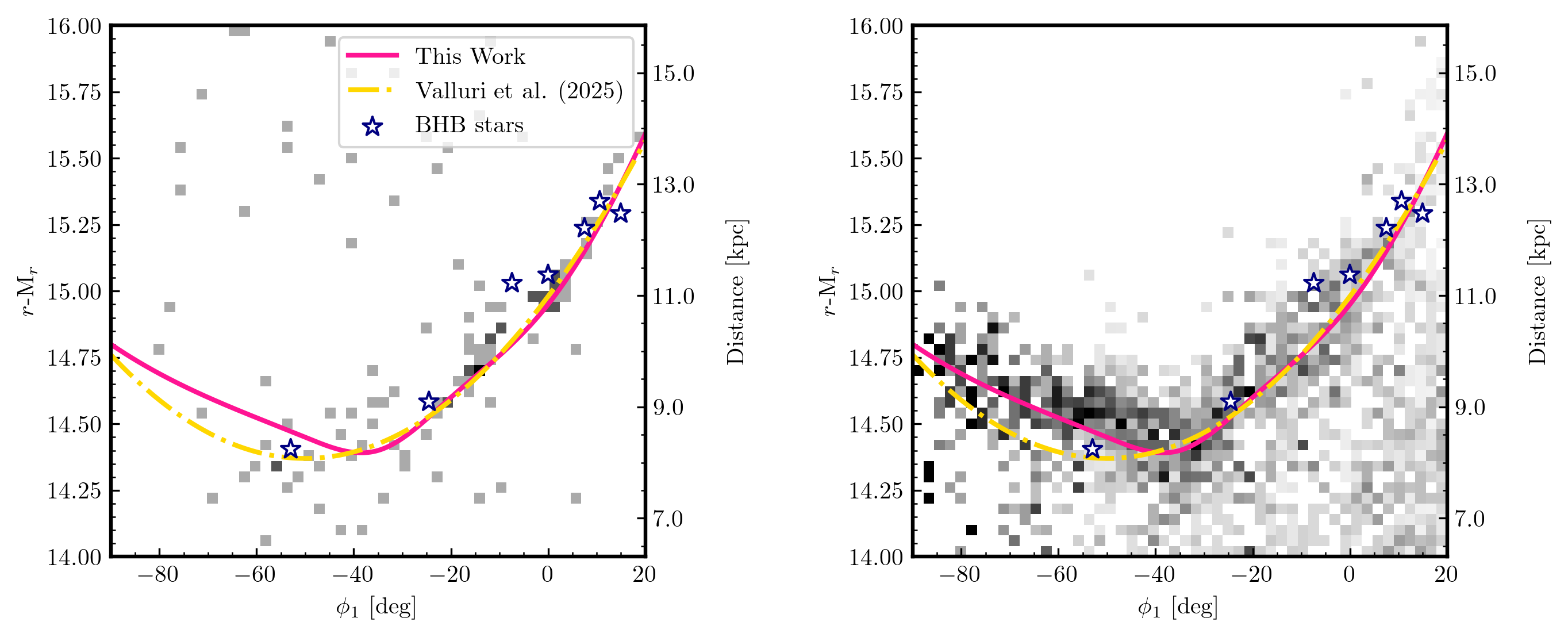}
    \caption{\textit{Left}: Distance moduli of DESI subgiant stars with selections $M_r < 3.69$, $|\delta\phi_2|<2^\circ$, $|\delta\mu_{\phi_1}|<1$~\masyr, $|\delta\mu_{\phi_2}|<1$~\masyr\ and $|\delta V_{\rm{GSR}}|<30$~\kms. \textit{Right}: Distance moduli of {\it Gaia} dwarf stars with selections $M_r > 3.69$, $|\delta\phi_2|<2^\circ$, $|\delta\mu_{\phi_1}|<1$~\masyr\ and $|\delta\mu_{\phi_2}|<1$~\masyr. Overlaid in pink is the improved distance gradient used in this work, obtained by including the dwarf stars. The yellow line depicts the distance gradient from \cite{valluri_2025}. Blue stars mark the locations of BHB stars along the stream.}
    \label{fig:dist-track}
\end{figure*}

\begin{figure*}
    \centering
    \includegraphics[width=\linewidth]{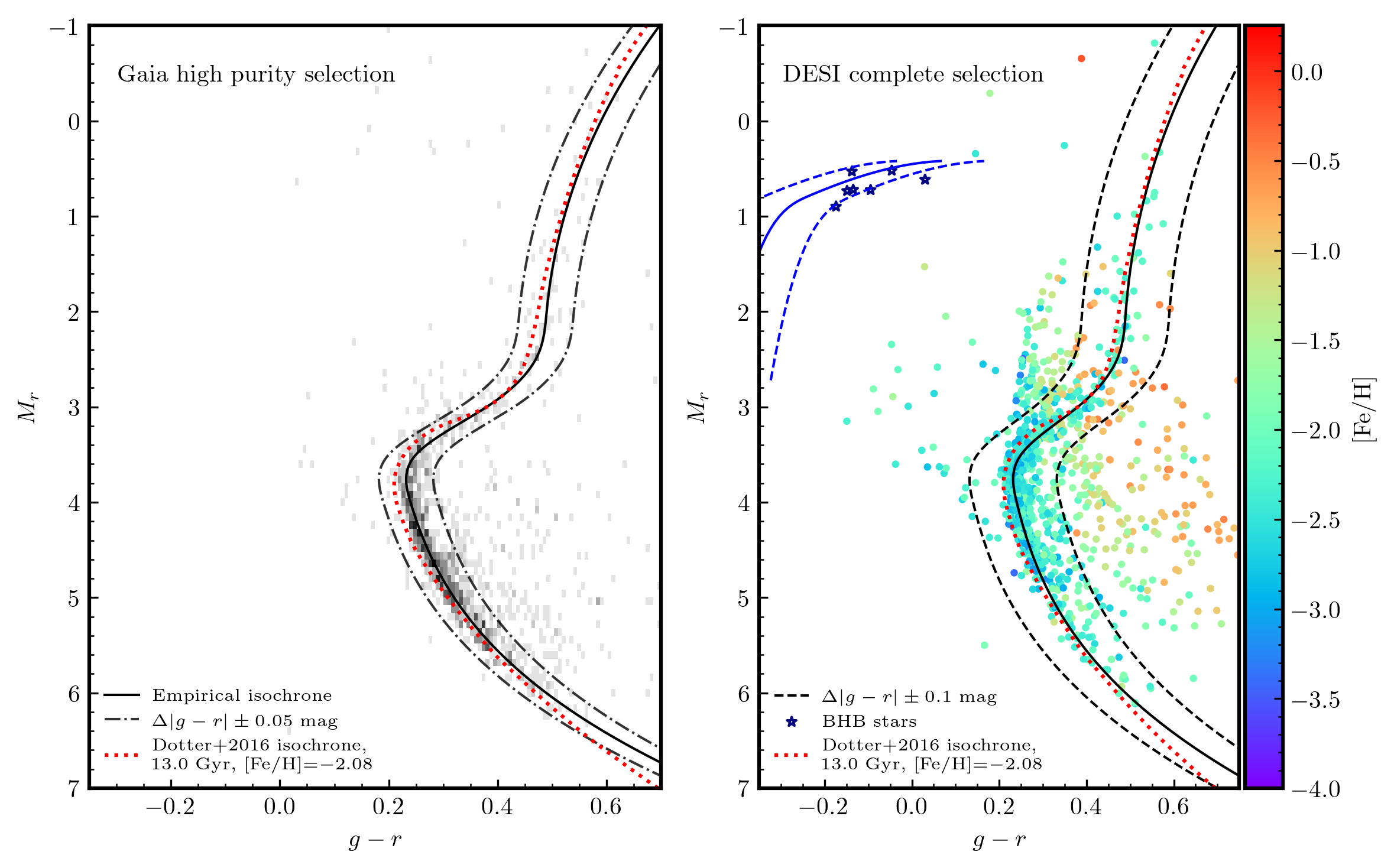}
    \caption{\textit{Left}: 2D histogram of color absolute magnitude of high purity {\it Gaia} stars selected by $|\delta\phi_2|<0.5^\circ$, $|\delta\mu_{\phi_1}|<0.5$~\masyr\ and $|\delta\mu_{\phi_2}|<0.5$~\masyr. Overlaid with a solid black line is the empirical isochrone that passes through the GD-1 population. Dashed dotted lines on either side of the isochrone are at $\delta|g-r| = 0.05$ and stars within these dotted lines are included in our selection of {\it Gaia} high purity stars, i.e. {\it Gaia} high purity selection. The dotted red line marks the track of the Dotter isochrone \citep{Dotter_2016} with an age of 13~Gyr and \feh~$=-2.08$, not used in this work but as a comparison to show how it does not follow the data (and therefore an empirical isochrone track is built instead). \textit{Right}: Color absolute magnitude of DESI stars in the GD-1 region with de-reddened magnitudes from DECam Legacy Survey photometry. Stars are selected with $|\delta\phi_2|<6^\circ$, $|\delta\mu_{\phi_1}|<2$~\masyr, $|\delta\mu_{\phi_2}|<2$~\masyr, and $|\delta V_{\textrm{GSR}}|<30$~\kms\ and points are colored by \feh\ values obtained with the DESI RVS pipeline. Dashed lines on either side of the isochrone are at $\delta|g-r| = 0.1$ and stars within these dashed lines are included in our complete selection of DESI GD-1 member stars, i.e. DESI complete selection. Blue stars mark the location of BHB stars, identified based on their position on the CMD relative to the BHB track (blue line).}
    \label{fig:cmd}
\end{figure*}

\begin{figure*}
    \centering
    \includegraphics[width=0.49\linewidth]{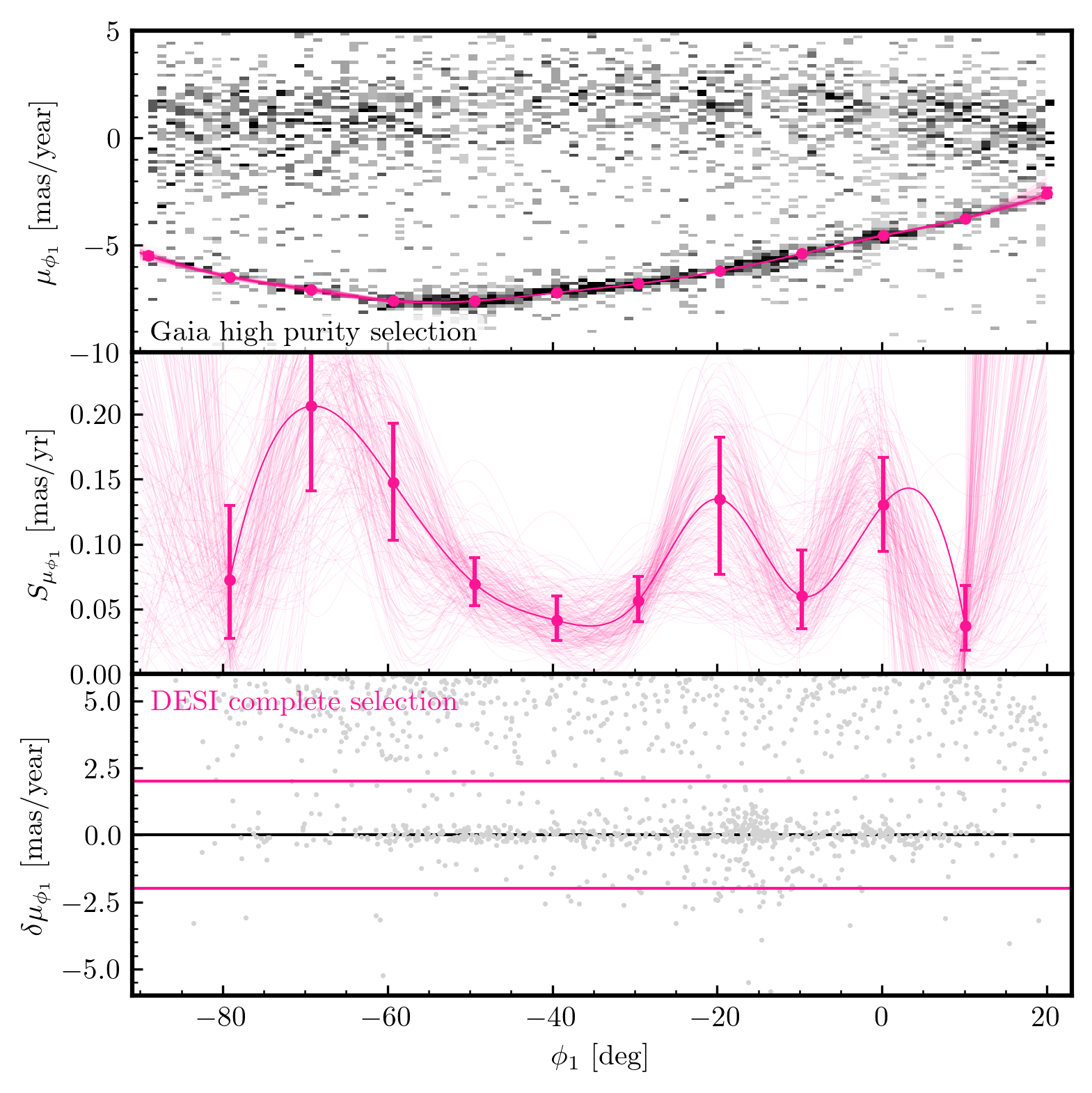}
    \includegraphics[width=0.49\linewidth]{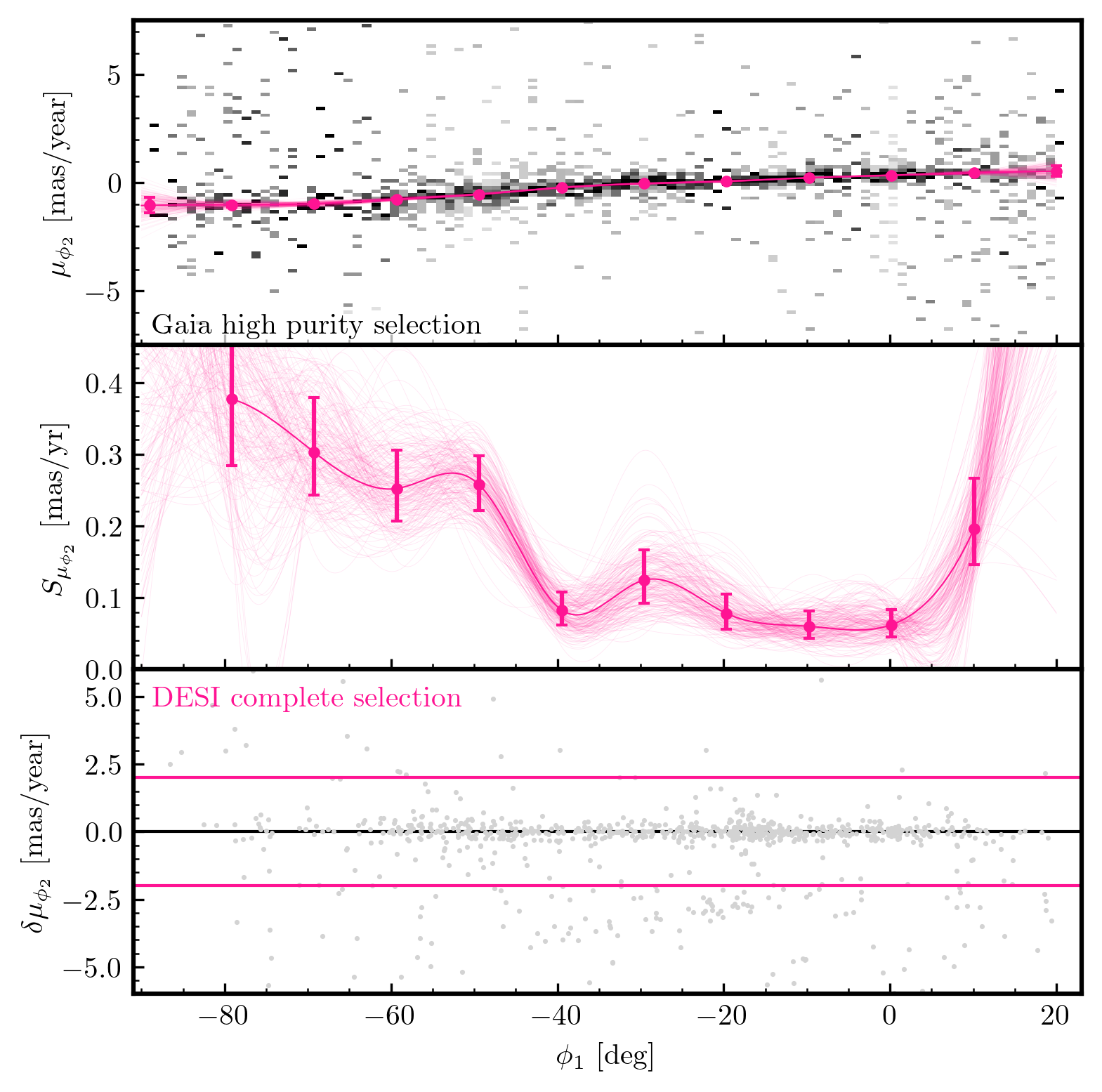}
    \includegraphics[width=0.49\linewidth]{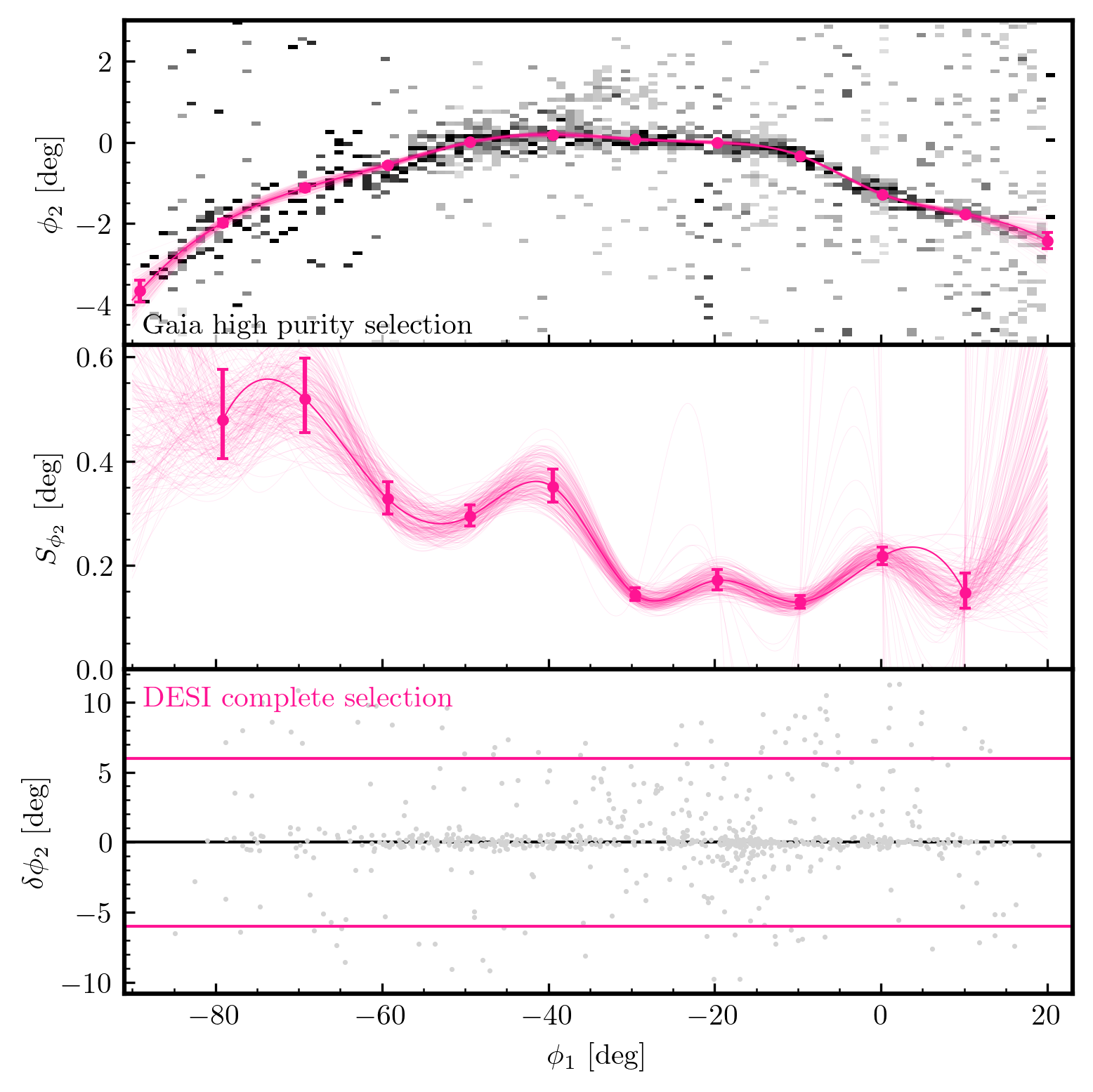}
    \includegraphics[width=0.49\linewidth]{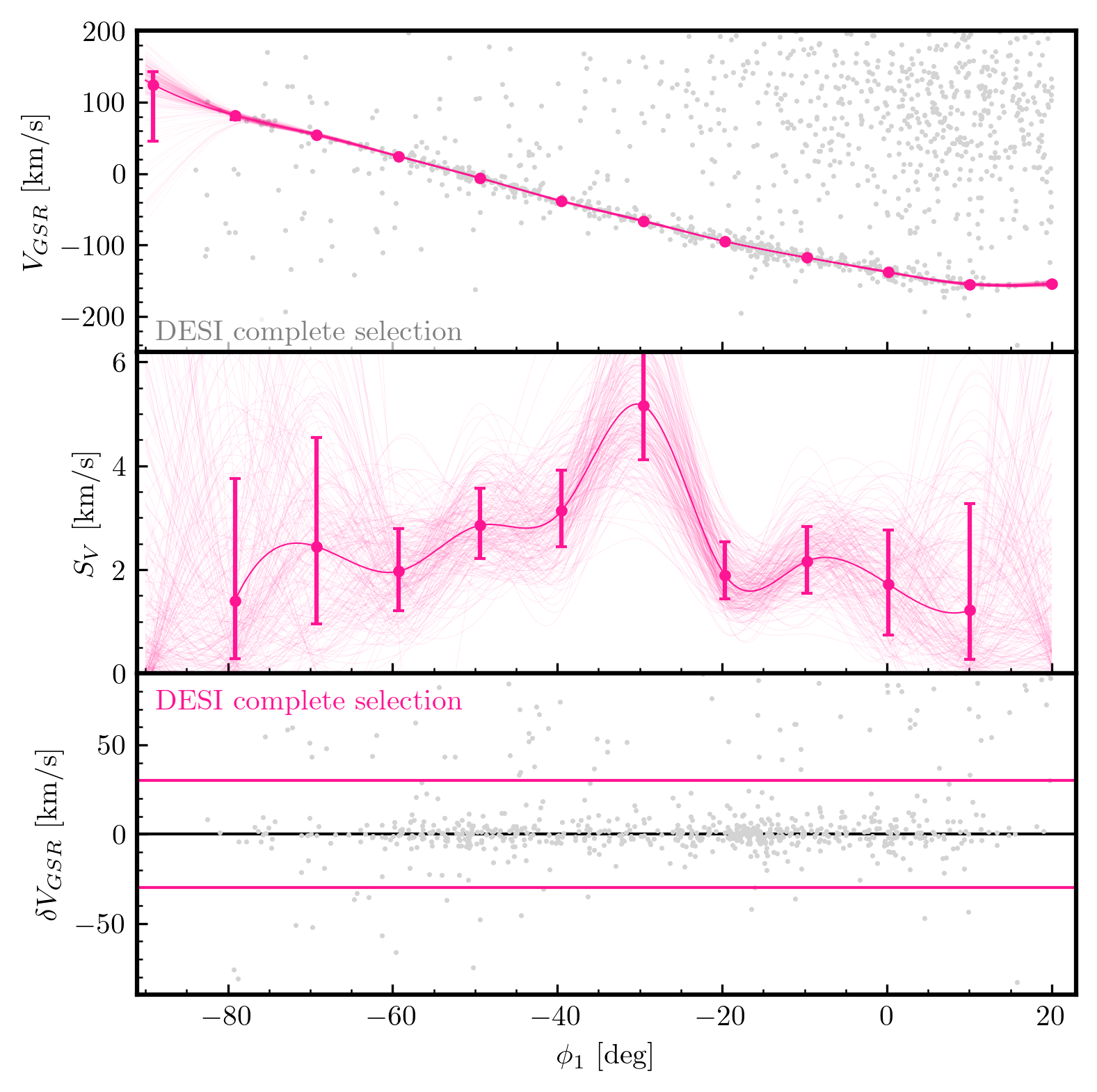}
    \caption{{\it Gaia} DR3 and DESI observations in the GD-1 region used to derive stream tracks in \pmone\ (top left set of plots), \pmtwo\ (top right), \phitwo\ (bottom left) and \vgsr\ (bottom right). \textit{Top panels}: the stars used to derive the stream track ({\it Gaia} DR3, high purity selection for \pmone, \pmtwo\ and \phitwo, and DESI DR2 with complete selection for \vgsr). Pink points depict the node locations obtained from the \texttt{Stan} spline Gaussian mixture model. Overplotted in pink is the stream track which is a spline passing through the nodes. \textit{Middle panels}: the dispersion of each parameter at each node location with samples from the posterior distribution overplotted in pink. Errorbars mark the 16\% and 84\% percentile of the posterior samples. \textit{Bottom panels}: Grey points depict the DESI data of each star with the stream track subtracted. Pink solid lines mark the cut used to obtain stream members with the DESI complete selection.}
    \label{fig:stan-results}
\end{figure*}

\subsection{Initial Selections}\label{sec:initial_sel}

We begin our identification of GD-1 members with a spatial selection in the GD-1 region by limiting our search to stars within $|\phi_2| < 10^\circ$ and $-90^\circ < \phi_1 < 20^\circ$. Our initial DESI quality cuts are \texttt{RR\_SPECTYPE == STAR} which ensures the spectroscopic type of the source is a star as classified by \texttt{REDROCK}, \texttt{RVS\_WARN == 0} to ensure there are no warnings  during the RVS fit, and \texttt{PRIMARY == TRUE}, which chooses the observation with the largest signal-to-noise, if a star has been observed more than once as a part of different surveys and programs. We limit the selection to stars with a radial velocity error of $<10$\,\kms. We also limit the selection to stars with \feh$>-3.99$ because the metallicity boundary is at \feh$=-4$ and stars found at the metal-poor boundary of the catalog are often the result of poor spectroscopic fits. 

We also impose a {\it Gaia} parallax cut such that

\begin{equation}
\begin{gathered}
\left|1 / d\left(\phi_1\right)-\varpi\right| < \varpi_\mathrm{sys}+2 \varpi_{e r r},
\end{gathered}
\end{equation}

\noindent where $\varpi$ is the parallax, $d\left(\phi_1\right)$ is the distance to the stream at a given value of $\phi_1$, $\varpi_{e r r}$ is the uncertainty on the {\it Gaia} parallax for each star, and $\varpi_\mathrm{sys}$ = 0.05\,mas is the systematic error on the parallax \citep{Lindegren_2021b}. For the initial parallax selection, the distances were obtained using the distance track from \cite{valluri_2025}, but the final member list of DESI members uses the distance track in Section \ref{sec:dist-track} to obtain the distance of each star for this parallax selection.

In order to derive new kinematic stream tracks, we make an initial cut to the sample of stars using the proper motion and \phitwo\ tracks as a function of \phione\ from \cite{valluri_2025}. We subtract each track to compute \dpmone, \dpmtwo\ and \dphitwo, relative to their respective stream tracks. We also include a CMD selection, which we will further describe in Section \ref{sec:cmd}. We define this initial strict cut of member stars as a ``{\it Gaia} high purity" selection given by:

\begin{itemize}
\setlength{\itemsep}{1pt}
\setlength{\parskip}{1pt}
\setlength{\parsep}{1pt}
\item $-90^\circ < \phi_1 < 20^\circ$
\item $|\delta\mu_{\phi_1}|<0.5$~\masyr\
\item  $|\delta\mu_{\phi_2}|<0.5$~\masyr\
\item  $|\delta\phi_2|<0.5^\circ$
\item $\Delta|g-r|<0.05.$
\end{itemize}

\noindent There are 916 {\it Gaia} stars that meet all of these high purity selections. We note that this is a selection using {\it Gaia} DR3 and DESI Legacy Survey data, i.e. no DESI spectroscopic data is used. Using an iterative approach, we start with these selected stars to then obtain improved tracks described in Section \ref{sec:track} and then use these new tracks to determine the final member list.

\subsection{The distance stream track}\label{sec:dist-track}

The distance to the GD-1 stream varies along \phione. To calibrate the distance gradient along the stream, we use the subgiant and dwarf stars. We separate the two populations with a magnitude cut at $M_r = 3.69$, and the subgiants and dwarfs from DESI and {\it Gaia} DR3 are plotted in the left and right panels of Figure \ref{fig:dist-track}, respectively. 
Both panels include proper motion selections of $|\delta\mu_{\phi_{1,2}}|<1$~\masyr\ and $|\delta\phi_2|<2^\circ$. For the subgiant sample (left panel), we additionally require $|\delta V_{\rm GSR}|<30$~\kms, where $\delta V_{\rm GSR}$ is the radial velocity deviation from the \vgsr\ stream track (Section~\ref{sec:rv-sel}), initially adopted from \citet{valluri_2025} and subsequently updated through the iterative procedure described in Section~\ref{sec:track}. This additional velocity cut is necessary because the subgiant sample is smaller and more susceptible to foreground contamination, and the DESI radial velocities provide the extra discriminating power needed to obtain a clean selection.

We define an empirical distance track by hand-picking node locations along $\phi_1$ that trace the peak density of the subgiant and dwarf sequences in distance moduli, and interpolate between the nodes using a cubic spline (shown in Figure~\ref{fig:dist-track}), with the node values presented in Table~\ref{tab:distance_track_nodes}.
This new distance track is then used to obtain the new CMD selection of stars, described in Section \ref{sec:cmd}. The primary difference between the track derived in this work and that used in \cite{valluri_2025} is the addition of the dwarf stars, and the improved track for the stream at $\phi_1 < -40^\circ$.

We also identify 7 blue horizontal branch (BHB) stars from their location on the CMD (see Figure \ref{fig:cmd}). A catalog of GD-1 BHB member stars is available in Table \ref{tab:bhb-members}. We obtain their distance using the relation from \cite{bystrom_2025}

\begin{equation}
\begin{aligned} 
    M_g &= 0.566 - 0.392(g-r)+2.729(g-r)^2 \\ 
    &+ 29.1128(g-r)^3 + 113.569(g-r)^4
\end{aligned}
\end{equation}
The BHB stars identified from the CMD closely follow the distance track, providing an independent check on our distance estimates. 

\begin{table}
\caption{Nodes of the GD-1 distance track.}
\label{tab:distance_track_nodes}
\begin{tabular}{ccc}
\toprule
$\phi_1$ [deg] & $m-M$ & $d$ [kpc] \\
\midrule
-90 & 14.80 & 9.12 \\
-70 & 14.60 & 8.32 \\
-50 & 14.45 & 7.76 \\
-35 & 14.40 & 7.59 \\
-20 & 14.60 & 8.32 \\
0 & 14.95 & 9.77 \\
25 & 15.80 & 14.45 \\
40 & 16.50 & 19.95 \\
\bottomrule
\end{tabular}
\end{table}

\subsection{Color-Magnitude Diagram Selection} \label{sec:cmd}
With the updated distance track, we next define a color-magnitude selection for GD-1 members. The CMD distribution is not well matched by a theoretical isochrone, so we construct an empirical CMD track that traces the observed stream population (see Figure~\ref{fig:cmd}). To do this, we hand-pick node locations along the range of $M_r$ values on the 2D histogram of the CMD, selecting the $g-r$ value that maximizes the stellar density at each node. This track is derived from the {\it Gaia} stars that satisfy our high-purity kinematic selection (Section~\ref{sec:initial_sel}), and the node locations are presented in Table~\ref{tab:cmd}. 

For the {\it Gaia} high-purity sample (Section~\ref{sec:initial_sel}), we adopt a tighter cut of $|\delta(g-r)| < 0.05$ (left panel of Figure~\ref{fig:cmd}) to ensure a clean sample for deriving the stream tracks, while for the full DESI sample we use a broader cut of $|\delta(g-r)| < 0.1$ (right panel of Figure~\ref{fig:cmd}; see Section~\ref{sec:mem_sel_summary} for the complete DESI selection criteria) to maximize completeness for the subsequent analysis. An isochrone cut of $\pm 0.1$ in $|g-r|$ around this empirical track has been shown in mock stream modeling to capture 80\% of stream stars while minimizing background contamination \citep{Holm-Hansen_2025}. The resulting empirical track closely follows the low-metallicity stellar population of the GD-1 stream.

\begin{table}[h!]
\centering
\begin{tabular}{rr}
\hline
$M_r$ & $g-r$ \\
\hline
$-2$   & 0.85 \\
0    & 0.59 \\
2    & 0.49 \\
2.5  & 0.47 \\
3    & 0.38 \\
3.5  & 0.25 \\
4    & 0.24 \\
5    & 0.32 \\
7    & 0.80 \\
\hline
\end{tabular}
\caption{Nodes of the CMD isochrone.}
\label{tab:cmd}
\end{table}

\subsection{Stream track derivation}\label{sec:track}

To perform selections in \pmone, \pmtwo, $\phi_2$ and \vgsr\ we fit empirical stream tracks and perform cuts around them. These tracks are used to isolate likely stream members from the broader Milky Way foreground/background stellar population. The stream tracks in \pmone, \pmtwo, and $\phi_2$ from \citet{valluri_2025} were determined by manually selecting node locations and fitting splines through them, without a statistical model to simultaneously account for the background population or to estimate the intrinsic stream dispersion. Only the \vgsr\ track in \citet{valluri_2025} was derived using a probabilistic mixture model. In this work, we derive all four tracks (\pmone, \pmtwo, $\phi_2$, and \vgsr) using a unified probabilistic framework based on a spline Gaussian mixture model (described below), which self-consistently separates the stream from the background and provides robust estimates of both the stream track and its dispersion as a function of \phione. 

To ensure uniform sampling along the length of the stream, we use stars from {\it Gaia} DR3 for the \pmone, \pmtwo, and $\phi_2$ tracks rather than the DESI sample, which has uneven coverage along $\phi_1$ (see Figure \ref{fig:completeness}). 

The derivation of our new tracks follows an iterative approach, beginning with the preliminary stream tracks from \citet{valluri_2025} as an initial starting point. We select {\it Gaia} stars using a CMD selection as described in Section~\ref{sec:cmd}, followed by spatial and kinematic cuts around these initial tracks -- specifically, stars within $\pm0.5$~\masyr\ in proper motion and within $\pm0.5^\circ$ from the \phitwo\ track (i.e.\ the {\it Gaia} high purity selection). We then fit new tracks using the probabilistic model described below and update the selections accordingly in the refit, as detailed in Section~\ref{sec:pm_sel}.

\subsubsection{Proper motion selection}\label{sec:pm_sel}

To refine our kinematic selections, we derive updated stream tracks for the proper motion components \pmone\ and \pmtwo\ using a probabilistic model. We use a truncated Gaussian mixture model, modified from that used in \cite{Koposov_2023}. We construct a model of the stream proper motions as a function of \phione. We model the stream component as a truncated Gaussian mixture where the center and dispersion of the Gaussian are allowed to vary with \phione\ and are represented by a spline with 12 evenly spaced nodes along the length of the stream. A second Gaussian represents the background contamination where the mean and dispersion are fitted parameters, and the mean can vary linearly along $\phi_1$. The likelihood function for this model is given by 

\begin{equation}\label{eq:stan}
\begin{split}
P\left(x \mid \phi_1\right) = & \left(1 - f\left(\phi_1\right)\right) \mathcal{N}\left(x \mid x_{b g} + d_{b g} \phi_1, \mathcal{S}_{b g}\right) \\
& + f\left(\phi_1\right) \mathcal{N}\left(x \mid \mathcal{X}\left(\phi_1\right), \mathcal{S}_{\mathcal{X}}\left(\phi_1\right)\right).
\end{split}
\end{equation}

\noindent where $f(\phi_1)$ is the stream fraction at each value of \phione, $x_{bg}$ and $d_{bg}$ define the background mean as a linear function of \phione, $\mathcal{S}_{bg}$ is the background dispersion, and $\mathcal{X}(\phi_1)$ and $\mathcal{S_X}(\phi_1)$ are the stream mean and dispersion, respectively.

We implement the model using the \texttt{Stan} programming language \citep{Carpenter_2017} and sample it using the \texttt{CMDstanpy}\footnote{\url{https://mc-stan.org/cmdstanpy/}} package. The splines were implemented using the \texttt{stan-splines} package \citep{Koposov_2023, Koposov_2024-stan}, following the approach described in the supplementary material of \cite{Koposov_2023}. We account for the proper motion errors by adding them in quadrature to the intrinsic proper motion dispersions when computing the likelihoods of individual stars.

For the spline-based Gaussian mixture model used to derive proper motion tracks, we adopt weakly informative priors on all parameters. The spline node values that define the proper motion mean as a function of \phione\ are assigned normal priors centered near the expected stream motion. Similarly, the dispersions are also assigned normal priors with large widths. The priors on all the background model parameters are also assigned normal distributions.

We then perform an iterative approach to obtain the new proper motion tracks. To begin, we fit an initial \pmone\ track after we select stars in CMD, $\phi_2$, and \pmtwo\ using the tracks from \cite{valluri_2025}. Similarly, we fit a new \pmtwo\ track by selecting stars in CMD, \phitwo\ and \pmone. Next, we apply the new distance track from this work (see Section \ref{sec:dist-track}) to recalculate the reflex correction for the proper motion values. With these updated values, we use the new \pmone\ track to select stars in \pmone\ and fit the final \pmtwo\ track. Finally, we reapply the selection with this updated \pmtwo\ track to fit our final \pmone\ track. The node locations of the final proper motion tracks are given in Table \ref{tab:Stan_tracks}. 

Figure \ref{fig:stan-results} depicts the results of the final \texttt{Stan} model for both \pmone\ and \pmtwo. The top panel shows a 2D histogram of the {\it Gaia} stars used to obtain the track. Overlaid are the nodes and the cubic spline through these nodes which we define as the track. The middle panel shows the dispersion along the stream in \masyr. The bottom panel shows the DESI stars that contain the CMD selection as well as the \phitwo\ and \vgsr\ selections (see Sections \ref{sec:phi2-sel} and \ref{sec:rv-sel}).

\begin{table*}
    \centering
    \begin{tabular}{rrrrrrrrr}
\toprule
$\phi_1$ & $\phi_2$ & $S_{\phi_2}$ & $\mu_{\phi_1}$ & $S_{\mu_{\phi_1}}$ & $\mu_{\phi_2}$ & $S_{\mu_{\phi_2}}$ & $V_{\mathrm{GSR}}$ & $S_{V_{\mathrm{GSR}}}$ \\
$\mathrm{[deg]}$ & $\mathrm{[deg]}$ & $\mathrm{[deg]}$ & $\mathrm{[mas~yr^{-1}]}$ & $\mathrm{[mas~yr^{-1}]}$ & $\mathrm{[mas~yr^{-1}]}$ & $\mathrm{[mas~yr^{-1}]}$ & $\mathrm{[km~s^{-1}]}$ & $\mathrm{[km~s^{-1}]}$ \\ \midrule
\midrule
$-89.06$ & $-3.66^{+0.27}_{-0.27}$ & $0.56^{+0.34}_{-0.19}$ & $-5.47^{+0.13}_{-0.15}$ & $0.20^{+0.46}_{-0.15}$ & $-1.04^{+0.38}_{-0.34}$ & $0.56^{+0.84}_{-0.27}$ & $124.79^{+17.79}_{-79.17}$ & $0.98^{+3.81}_{-0.78}$ \\
$-79.15$ & $-1.99^{+0.09}_{-0.09}$ & $0.48^{+0.10}_{-0.07}$ & $-6.47^{+0.06}_{-0.06}$ & $0.07^{+0.06}_{-0.05}$ & $-1.02^{+0.10}_{-0.10}$ & $0.38^{+0.10}_{-0.09}$ & $81.58^{+2.17}_{-5.57}$ & $1.40^{+2.35}_{-1.12}$ \\
$-69.23$ & $-1.12^{+0.08}_{-0.08}$ & $0.52^{+0.08}_{-0.06}$ & $-7.07^{+0.07}_{-0.07}$ & $0.21^{+0.07}_{-0.07}$ & $-0.96^{+0.07}_{-0.07}$ & $0.30^{+0.08}_{-0.06}$ & $54.53^{+1.44}_{-1.47}$ & $2.44^{+2.11}_{-1.49}$ \\
$-59.32$ & $-0.56^{+0.04}_{-0.04}$ & $0.33^{+0.03}_{-0.03}$ & $-7.60^{+0.04}_{-0.04}$ & $0.15^{+0.05}_{-0.04}$ & $-0.76^{+0.04}_{-0.04}$ & $0.25^{+0.05}_{-0.05}$ & $24.36^{+0.68}_{-0.66}$ & $1.97^{+0.82}_{-0.76}$ \\
$-49.40$ & $0.00^{+0.03}_{-0.03}$ & $0.29^{+0.02}_{-0.02}$ & $-7.60^{+0.02}_{-0.02}$ & $0.07^{+0.02}_{-0.02}$ & $-0.53^{+0.03}_{-0.03}$ & $0.26^{+0.04}_{-0.04}$ & $-6.21^{+0.59}_{-0.59}$ & $2.85^{+0.72}_{-0.64}$ \\
$-39.48$ & $0.19^{+0.03}_{-0.03}$ & $0.35^{+0.03}_{-0.03}$ & $-7.21^{+0.02}_{-0.02}$ & $0.04^{+0.02}_{-0.02}$ & $-0.22^{+0.02}_{-0.02}$ & $0.08^{+0.02}_{-0.02}$ & $-38.53^{+0.75}_{-0.75}$ & $3.14^{+0.78}_{-0.69}$ \\
$-29.57$ & $0.08^{+0.02}_{-0.02}$ & $0.14^{+0.01}_{-0.01}$ & $-6.79^{+0.02}_{-0.02}$ & $0.06^{+0.02}_{-0.02}$ & $-0.03^{+0.02}_{-0.02}$ & $0.12^{+0.04}_{-0.03}$ & $-66.31^{+0.88}_{-0.87}$ & $5.16^{+1.25}_{-1.05}$ \\
$-19.65$ & $-0.01^{+0.02}_{-0.02}$ & $0.17^{+0.02}_{-0.02}$ & $-6.21^{+0.02}_{-0.03}$ & $0.13^{+0.05}_{-0.06}$ & $0.07^{+0.02}_{-0.02}$ & $0.08^{+0.03}_{-0.02}$ & $-94.87^{+0.37}_{-0.36}$ & $1.90^{+0.64}_{-0.46}$ \\
$-9.74$ & $-0.33^{+0.02}_{-0.02}$ & $0.13^{+0.01}_{-0.01}$ & $-5.39^{+0.02}_{-0.02}$ & $0.06^{+0.04}_{-0.03}$ & $0.23^{+0.02}_{-0.02}$ & $0.06^{+0.02}_{-0.02}$ & $-117.47^{+0.49}_{-0.49}$ & $2.16^{+0.68}_{-0.62}$ \\
$0.18$ & $-1.28^{+0.02}_{-0.02}$ & $0.22^{+0.02}_{-0.02}$ & $-4.54^{+0.02}_{-0.03}$ & $0.13^{+0.04}_{-0.04}$ & $0.35^{+0.02}_{-0.02}$ & $0.06^{+0.02}_{-0.02}$ & $-137.98^{+0.63}_{-0.58}$ & $1.72^{+1.05}_{-0.98}$ \\
$10.09$ & $-1.77^{+0.04}_{-0.04}$ & $0.15^{+0.04}_{-0.03}$ & $-3.75^{+0.03}_{-0.03}$ & $0.04^{+0.03}_{-0.02}$ & $0.46^{+0.05}_{-0.05}$ & $0.20^{+0.07}_{-0.05}$ & $-154.72^{+0.99}_{-1.01}$ & $1.22^{+2.05}_{-0.94}$ \\
$20.01$ & $-2.43^{+0.21}_{-0.19}$ & $0.41^{+0.38}_{-0.19}$ & $-2.59^{+0.27}_{-0.17}$ & $0.60^{+1.94}_{-0.44}$ & $0.54^{+0.24}_{-0.24}$ & $0.98^{+0.54}_{-0.39}$ & $-154.25^{+1.95}_{-2.15}$ & $0.86^{+2.44}_{-0.65}$ \\
\bottomrule
\end{tabular}

    \caption{Stream track and dispersion track node locations. A machine-readable version of this table is available on Zenodo (see Data Availability.)}
    \label{tab:Stan_tracks}
\end{table*}

\subsubsection{$\phi_2$ selection}\label{sec:phi2-sel}
We use the same approach as described in Section \ref{sec:pm_sel} to obtain an improved stream track in \phitwo. We use the likelihood function defined in Equation \ref{eq:stan} to select {\it Gaia} stars within $\pm0.5$~\masyr\ of the final proper motion stream tracks derived in Section \ref{sec:pm_sel}. The same {\it Gaia} high purity CMD selection is applied as well. We then fit the spline to the stars that pass these cuts. The priors on model parameters are again normal distributions with large standard deviations. Figure \ref{fig:stan-results} depicts the results of this \texttt{Stan} model. The node locations of this \phitwo\ track are given in Table \ref{tab:Stan_tracks}. For our DESI member selection, we keep all stars within $|\delta \phi_2| < 6^\circ$ from the $\phi_2$ track.

\subsubsection{Radial velocity selection}\label{sec:rv-sel}
We use DESI radial velocities to obtain a new \vgsr\ track for GD-1, where \vgsr\ is the solar reflex-motion corrected line-of-sight velocity in the Galactic Standard of Rest frame. Using our newly derived proper motion and $\phi_2$ tracks described in Sections \ref{sec:pm_sel} and \ref{sec:phi2-sel}, we select stars with $|\delta\mu_{\phi_{1, 2}}|<2$\,\masyr\ and $|\delta\phi_2|<3^\circ$. We further apply our DESI complete CMD selection, a cut of \feh$<-1.5$, and the initial selections described in Section \ref{sec:initial_sel} to derive the \vgsr\ track. These cuts leave a total of \comment{1314} DESI stars prior to cuts in radial velocity. 

To derive the radial velocity track, we use the same model as described in Section \ref{sec:pm_sel}, using the likelihood function defined in Equation \ref{eq:stan}. The priors on model parameters are normal distributions with large standard deviations. Figure \ref{fig:stan-results} depicts the results of this mixture model. Finally, we select stars within $\pm30$~\kms\ from the derived \vgsr\ track as the GD-1 members.

\subsection{Summary of DESI complete selection}\label{sec:mem_sel_summary}

Using the tracks derived in Section \ref{sec:track}, we obtain \dvgsr, \dpmone, \dpmtwo, and \dphitwo\ for the member stars by subtracting the stream track from each star and these values are used to impose selections to the stars. To summarize, our initial GD-1 DESI membership selection meets the following cuts, on top of the primary cuts described in Section \ref{sec:initial_sel}, which we define as the ``DESI complete selection":

\begin{itemize}
    \setlength{\itemsep}{1pt}
    \setlength{\parskip}{1pt}
    \setlength{\parsep}{1pt}
    \item$-90^\circ < \phi_1 < 20^\circ$
    \item $V_{\mathrm{err}}< 10$~\kms
    \item $|\delta (g - r)|< 0.1$
    \item $|\delta\mu_{\phi_1}|< 2$~\masyr
    \item $|\delta\mu_{\phi_2}|< 2$~\masyr
    \item $|\delta\phi_2|< 6^\circ$
    \item $|\delta V_{\textrm{GSR}}|< 30$~\kms
    \item \feh\ $<-1.5$
\end{itemize}

\noindent We obtain \comment{679} DESI stars that meet these initial selections. Without including any \feh\ selection, there would be \comment{736} member candidates and without the \vgsr\ selection, this membership criteria would yield \comment{3355} member candidates.

\section{Characterizing the stream components}\label{sec:GMM}

With our new catalog of DESI GD-1 stars obtained using the membership selection described in Section \ref{sec:mem_sel_summary}, we then analyze the components of the stream, in particular we separate the thin stream and cocoon components of GD-1 as defined in \citet{valluri_2025}.

\subsection{Gaussian mixture model analysis}\label{sec:GMM-description}

To characterize the components of the stream, we model the observed phase-space properties using a Gaussian mixture model. We fit the data with two Gaussian components representing the ``thin stream" and the ``cocoon" along with a uniform background distribution to account for residual contamination not associated with GD-1. For this analysis, we restrict our sample of stars to within $-60^\circ < \phi_1 < 0^\circ$, corresponding to the portion of the stream most densely covered by DESI observations (see Figure \ref{fig:desi-releases}). Outside of this range, we have a less complete sample of member stars. With this $\phi_1$ cut, \comment{545} stars remain for this analysis.

\begin{table}[h]
\centering
\begin{tabular}{lll}
\hline
Parameter & Stream Prior & Cocoon Prior \\
\hline

$\overline{\delta V}$ & $[-10,\ 10]$ & $[-10,\ 10]$ \\
$\Sigma_{\delta V}$ & $(0,\ 10]$ & $(0,\ 20]$ \\

$\overline{\delta \mu_{\phi_1}}$ & $[-2,\ 2]$ & $[-2,\ 2]$ \\
$\log_{10}\Sigma_{\delta \mu_{\phi_1}}$ & $[-2,\ 2]$ & $[-2,\ 2]$ \\

$\overline{\delta \mu_{\phi_2}}$ & $[-2,\ 2]$ & $[-2,\ 2]$ \\
$\log_{10}\Sigma_{\delta \mu_{\phi_2}}$ & $[-2,\ 2]$ & $[-2,\ 2]$ \\

$\overline{\delta \phi_2}$ & $[-2,\ 2]$ & $[-2,\ 2]$ \\
$\log_{10}\Sigma_{\delta \phi_2}$ & $[-2,\ 2]$ & $[-2,\ 2]$ \\

$Q$ & $[0,\ 1]$ & $[0,\ 1]$ \\

\hline
\end{tabular}
\caption{Prior ranges used in the likelihood for the stream and cocoon components. The model also enforces that the stream is narrower than the cocoon in all dimensions. Velocity dispersion priors are applied in linear space, while the proper-motion and $\phi_2$ dispersion priors are applied in log space.}
\label{tab:priors}
\end{table}

The likelihood function for the mixture model is expressed as:

\begin{equation}
\begin{split}
    \mathcal{L} = & \prod_{i=1}^N Q_1 \mathcal{N}_1\left(x_i \mid \bar{x}_1, \Sigma_1\right) + Q_2 \mathcal{N}_2\left(x_i \mid \bar{x}_2, \Sigma_2\right) \\
    & + (1 - Q_1 - Q_2) \mathcal{U}(x_i)
\end{split}
\end{equation}

\begin{table*}[ht]
\centering
\caption{Gaussian Mixture Model parameter comparison between 1-component and 2-component fits.}
\label{tab:gmm_comparison}
\begin{tabular}{lrrr}
\toprule
& 1 Component & \multicolumn{2}{c}{2 Components} \\
\cmidrule(lr){2-2} \cmidrule(lr){3-4}
Parameters & Stream & Thin Stream & Cocoon \\
\midrule
\midrule
Number of members ($p>0.5$) & 467 & 421 & 185 \\
Number of expected members ($\sum p$) & 433 & 407 & 194 \\
Q & 0.70 ± 0.02 & 0.60 ± 0.02 & 0.30 ± 0.02 \\
$\overline{\delta V}$ [km/s] & --0.26 ± 0.22 & --0.29 ± 0.22 & 0.72 ± 0.62 \\
$\Sigma_{\delta V}$ [km s$^{-1}$] & 2.75 ± 0.31 & 2.49 ± 0.28 & 6.13 ± 0.75 \\
$\overline{\delta \mu_{\phi_1}}$ [km s$^{-1}$] & 1.35 ± 0.33 & 0.27 ± 0.33 & --5.78 ± 4.10 \\
$\Sigma_{\delta \mu_{\phi_1}}$ [km s$^{-1}$] & 5.43 ± 0.89 & 2.65 ± 0.43 & 41.36 ± 4.98 \\
$\overline{\delta \mu_{\phi_2}}$ [km s$^{-1}$] & 0.37 ± 0.28 & 0.39 ± 0.28 & --0.34 ± 0.52 \\
$\Sigma_{\delta \mu_{\phi_2}}$ [km s$^{-1}$] & 0.74 ± 0.35 & 0.68 ± 0.32 & 3.62 ± 0.65 \\
$\overline{\delta \phi_2}$ [deg] & 0.01 ± 0.01 & --0.01 ± 0.01 & 0.53 ± 0.19 \\
$\Sigma_{\delta \phi_2}$ [deg] & 0.37 ± 0.02 & 0.23 ± 0.01 & 2.18 ± 0.17 \\
\bottomrule
\end{tabular}
\end{table*}

\noindent where $x_i$ denotes the observed quantities with the stream track subtracted. These includes four variables: the line-of-sight velocity ($\delta$\vgsr), proper motions ($\delta$\pmone, $\delta$\pmtwo), and $\delta$\phitwo. The terms $Q_1$ and $Q_2$ represent the fractional contributions of the thin stream and cocoon components, respectively, while $\mathcal{N}$ denotes multivariate normal distributions parameterized by $\bar{x}_k$ and $\Sigma_k$ for each component $k$. The term  $\mathcal{U}(x_i)$ represents the uniform distribution used to model the background and foreground population.

The proper motion measurements are given in angular units (\masyr), but the velocity dispersion of the GD-1 stream is more naturally expressed in \kms, requiring a conversion that depends on distance as $v = 4.74 \, \mu \, d$, with $\mu$ in \masyr\ and $d$ in kpc. Since the distance to the stream varies along $\phi_1$, the same proper motion dispersion in \masyr\ corresponds to different transverse velocity dispersions in \kms\ at different positions along the stream. To account for this, we rescale the proper motion of each star to the distance at $\phi_1 = -30^\circ$, roughly the midpoint of the stream, which corresponds to a distance of 7.75~kpc from the distance track derived in Section~\ref{sec:dist-track}. This allows us to fit the stream with a single velocity dispersion in \kms\ across the full range of $\phi_1$, under the assumption that the intrinsic transverse velocity dispersion of GD-1 is constant along the length of the stream.

We initially estimate the model parameters with direct maximization of the likelihood function using \texttt{scipy.optimize}. We then use these estimates as the initial guess to generate the posterior distribution of the model parameters using the affine-invariant Markov Chain Monte Carlo (MCMC) ensemble sampler implemented in \texttt{emcee} \citep{Foreman_2013}. We adopt uniform priors for all parameters as detailed in Table \ref{tab:priors}. The posterior samples are generated with 64 walkers. We run 2000 steps and we discard the first 1000 steps as burn-in.

\begin{figure*}
    \centering
    \includegraphics[width=0.48\linewidth]{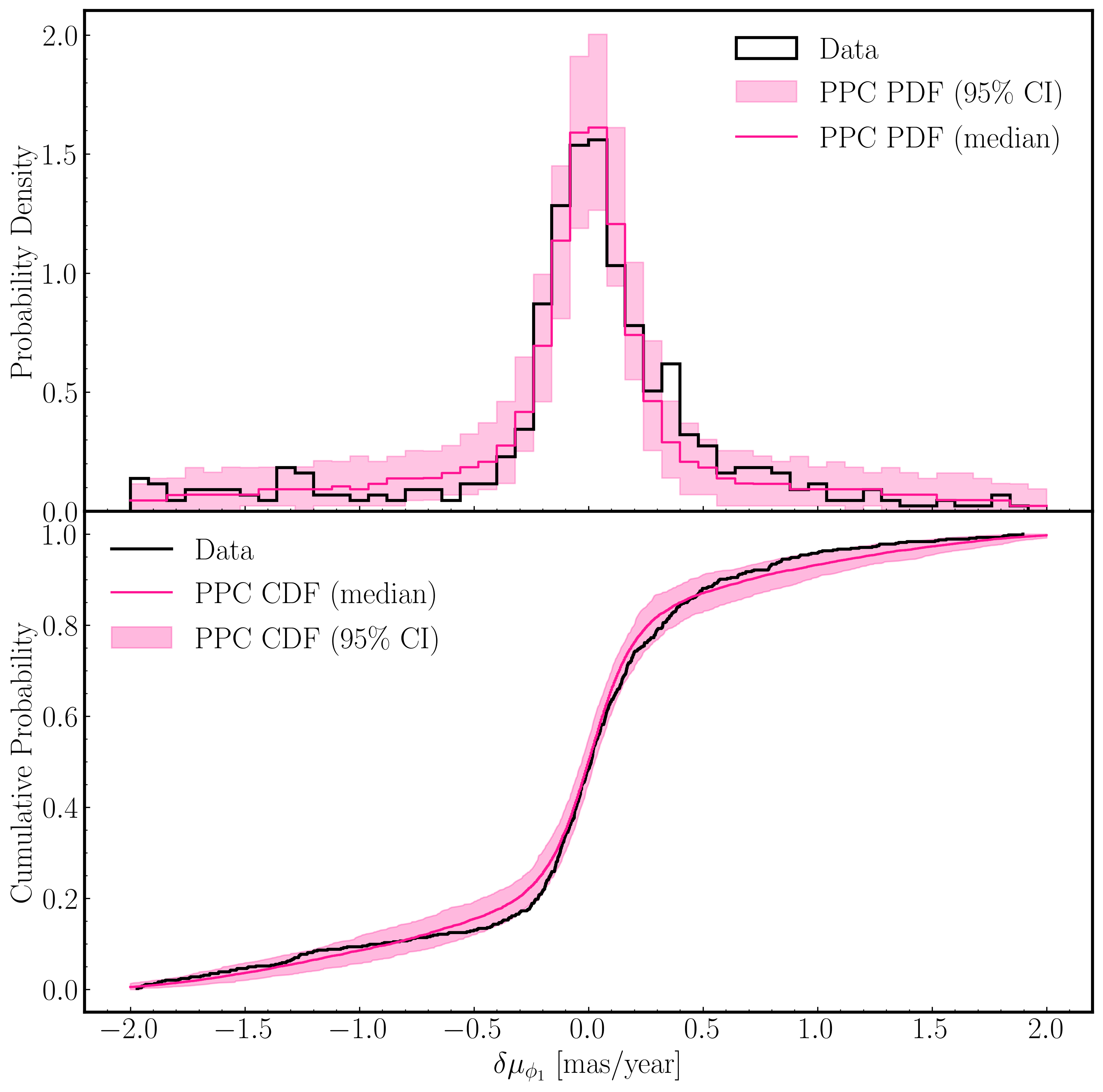}
    \includegraphics[width=0.48\linewidth]{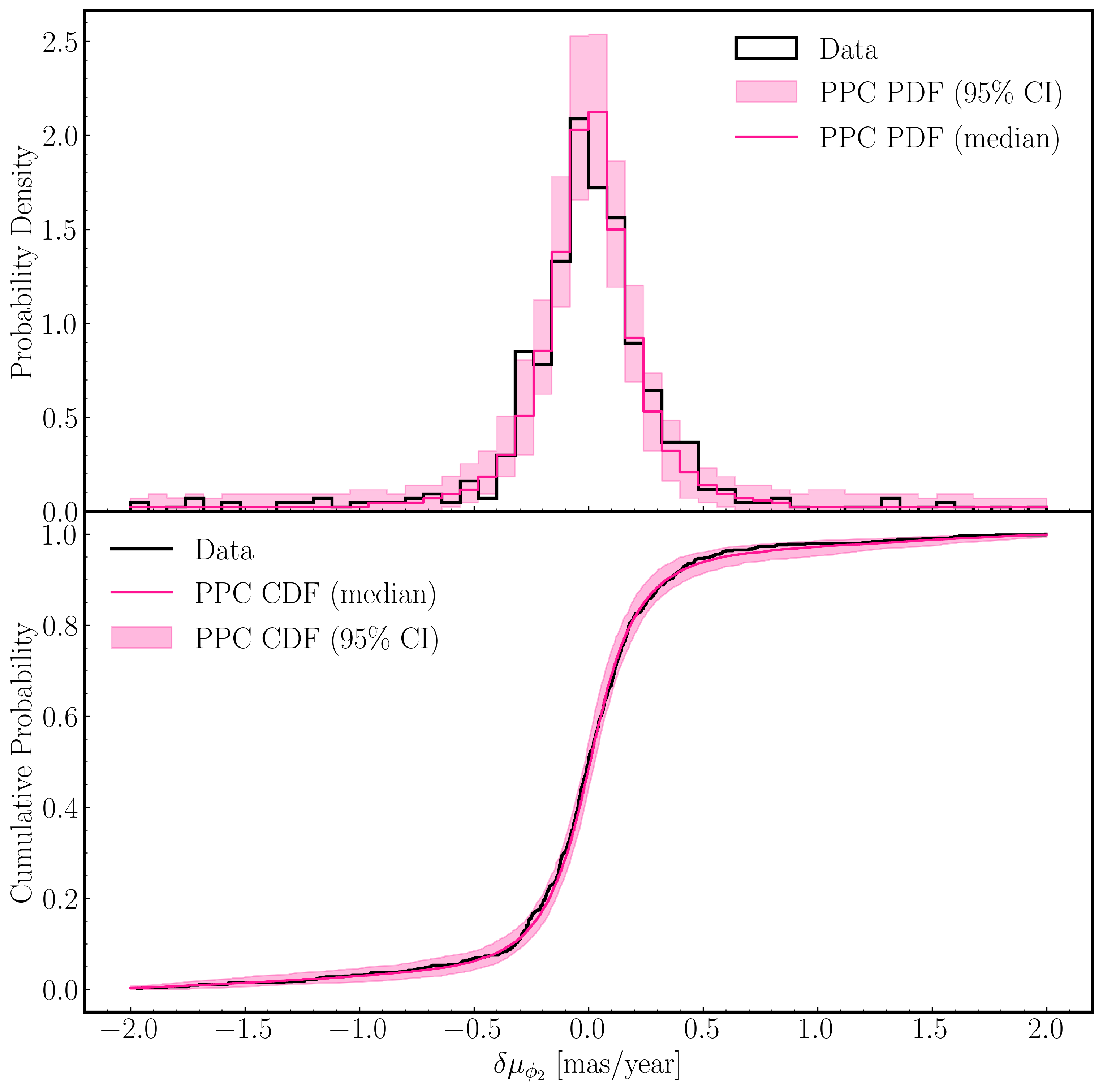}
     \includegraphics[width=0.48\linewidth]{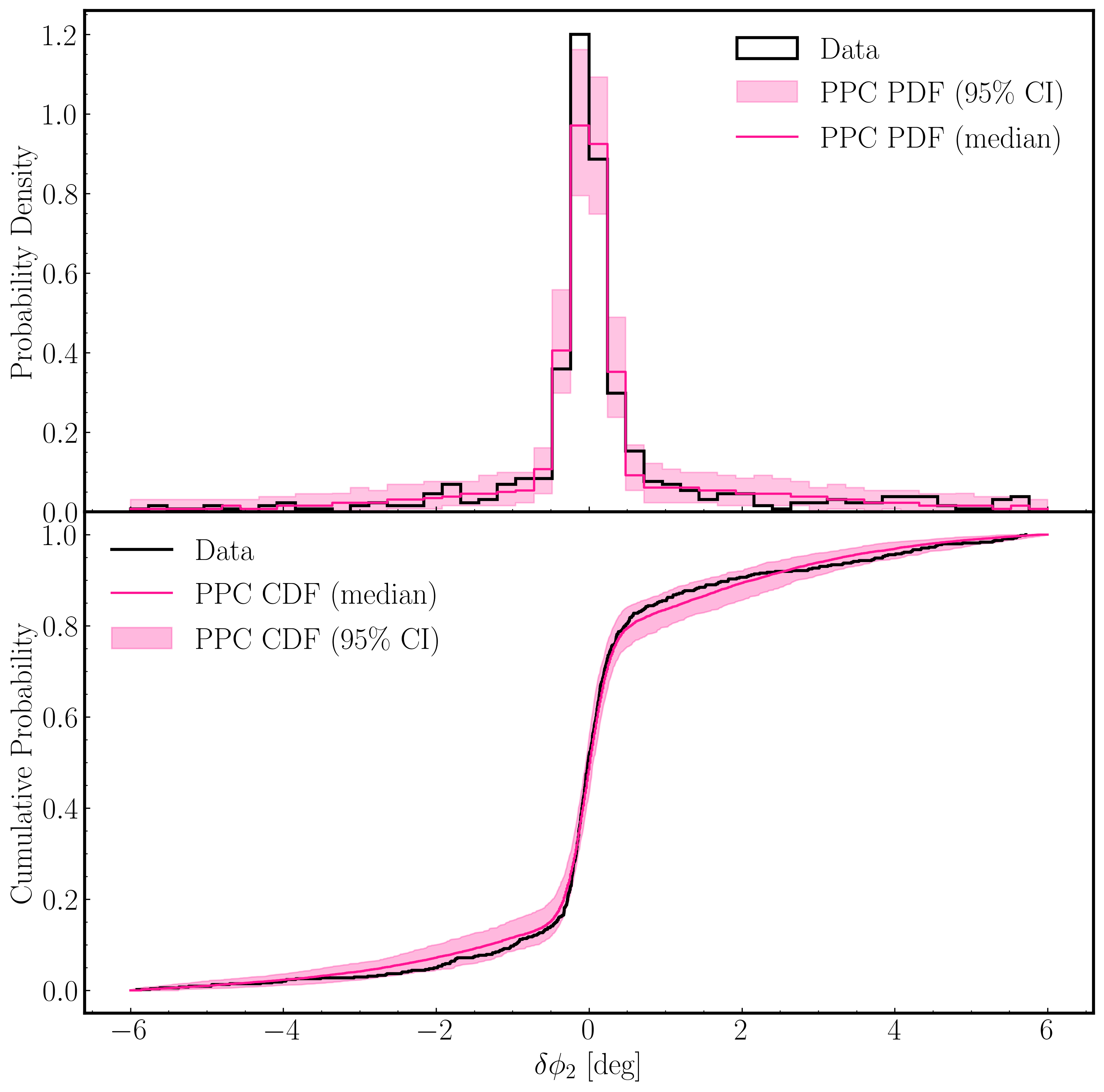}
     \includegraphics[width=0.48\linewidth]{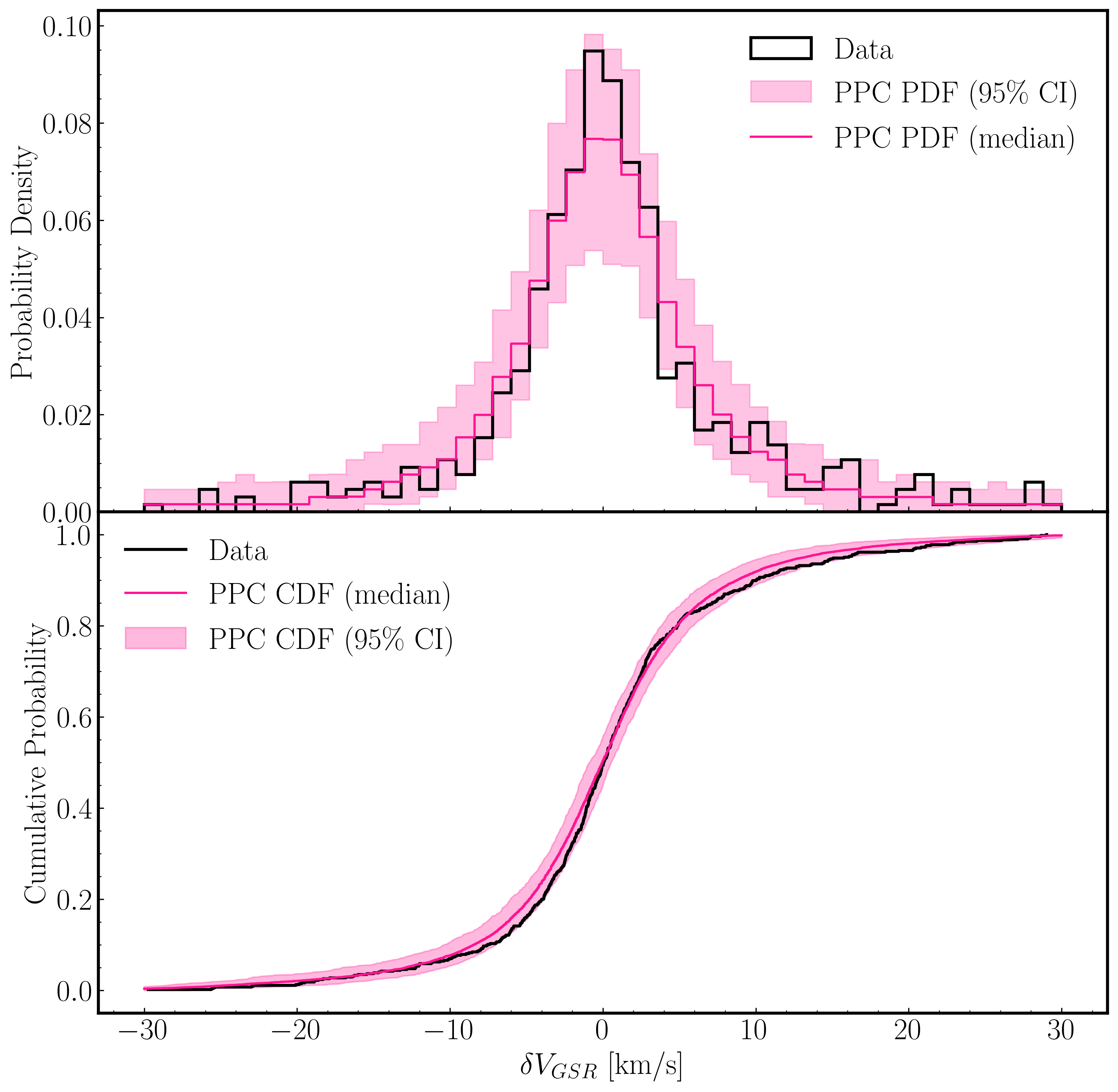}
    \caption{Results of the Gaussian mixture model used to characterize the stream and cocoon for each of the four stream properties: \dpmone\ (top left), \dpmtwo\ (top right), \dphitwo\ (bottom left), and \dvgsr\ (bottom right). \textit{Top panels}: histograms of the data used in the mixture model (black line) compared with the posterior predictive check (PPC) probability density functions (PDFs). The PPC distributions are obtained by drawing 200 samples from the posterior and generating mock datasets by sampling from the model by adding measurement uncertainties for each star. In each panel, the pink line shows the median PPC PDF, and the shaded pink region shows the 95\% credible interval (CI) of the model. \textit{Bottom panels}: cumulative distribution functions (CDFs) using the same color scheme as the top panels.}
    \label{fig:gmm-ppc}
\end{figure*}

 In Figure \ref{fig:gmm-ppc} we show the data compared with posterior predictive realizations of the model. We assess the robustness of the model with posterior predictive checks by comparing the cumulative distribution functions (CDFs) of the four observables generated by the model to the empirical CDFs. To do this, we draw 200 random realizations from the posterior distribution of model parameters, and for each realization we generate a mock dataset by sampling from the model and adding the measurement uncertainties for each star from the observations (e.g., {\it Gaia} uncertainties for proper motions, DESI uncertainties for velocities). The observed distributions lie within the 95\% credible interval of the posterior predictive samples, indicating that once measurement errors are accounted for, our three-component mixture model provides an adequate description of the intrinsic populations.

Table \ref{tab:gmm_comparison} summarizes the results from the Gaussian mixture model. We report the parameter estimates as the 50th percentile of the marginalized posterior distributions. We report symmetric uncertainties as the average of the upper and lower deviations from the median to the 84th and 16th percentiles, respectively. The fraction of stars in the thin stream is \comment{$0.60\pm0.02$} and the fraction in the cocoon is \comment{$0.30\pm0.02$} which leaves a background fraction of \comment{$0.10\pm0.04$}. For all parameters, the mean value is close to zero. This is expected because the stream track has been subtracted. The only exception is the \pmone\ mean, which will be discussed in Section \ref{sec:pm_disp}. The standard deviation of the \phitwo\ Gaussian fit to the thin and cocoon components are \comment{$(0.23\pm0.01)^\circ$} and \comment{$(2.18\pm0.17)^\circ$}, respectively. 
Assuming a mean distance to the stream of 7.75~kpc (the mean distance to the stream within $-60^\circ<\phi_1<0^\circ$), this corresponds to a physical width of \comment{$\approx78$~pc} for the thin stream and \comment{$\approx743$~pc} for the cocoon.

For the \vgsr, the cocoon component exhibits a broader dispersion \comment{($\Sigma_{\delta V} = 6.13\pm0.75$~\kms)} compared to the thin stream \comment{($\Sigma_{\delta V} = 2.49\pm0.28$~\kms)}. The typical measurement uncertainty is $\sim3$~\kms\ which means the thin stream's velocity dispersion is only marginally resolved, while the cocoon's is clearly broader. For the proper motion in the $\phi_1$ direction, the thin stream has a small dispersion (\comment{$2.65\pm0.43$~\kms}) while the cocoon shows a much larger dispersion (\comment{$41.36\pm4.98$~\kms}), which will be discussed further in Section \ref{sec:pm_disp}. A similar, though less pronounced, trend is seen in \pmtwo. The thin stream has a dispersion of \comment{$0.74\pm0.35$~\kms} and the cocoon has slightly broader dispersion of \comment{$3.62\pm0.65$~\kms}. 
However, these values should be interpreted with caution. The typical proper motion uncertainty of $\sim0.15$~\masyr\ corresponds to $\sim6$~\kms\, which is substantially larger than the fitted dispersions. Although our mixture model accounts for measurement uncertainties, when the intrinsic dispersion is much smaller than the measurement error, the inferred value is poorly constrained and may not be reliably resolved. We therefore do not draw physical conclusions from the \pmtwo\ dispersions.

\subsubsection{One Gaussian component model}

As a comparison with the two-Gaussian model, we also consider a model consisting of a single Gaussian to represent the stream component, along with a uniform component for the background and foreground stars. The results of this model are presented in the first column of Table \ref{tab:gmm_comparison}. With this model, we find a stream fraction of $\mathrm{Q}=\comment{0.70}\pm0.02$. We note that for the one component model, the values of the dispersion are typically between the values obtained from the two component model, although slightly closer to the thin stream value due to the higher fraction of stars in the thin stream. The values of the mean are also typically closer to zero for the one-component model.

To compare the selection of the two-Gaussian plus uniform model to the single-Gaussian plus uniform model, we use the Akaike Information Criterion \citep[AIC;][]{Akaike_1974} and the Bayesian Information Criterion \citep[BIC;][]{Schwarz_1978}, where lower values indicate a better fit. Although the two-Gaussian model includes nine additional parameters relative to the one-Gaussian model, both criteria strongly favor it, with the two-Gaussian model yielding $\Delta \mathrm{AIC} = \comment{-496}$ and $\Delta\mathrm{BIC} = \comment{-457}$ relative to the one-Gaussian model. These large negative differences indicate a decisive preference for the two-component description of the stream, supporting the decomposition into a thin stream and cocoon.

\subsection{Membership probability}

\begin{figure*}
    \centering
    \includegraphics[width=\linewidth]{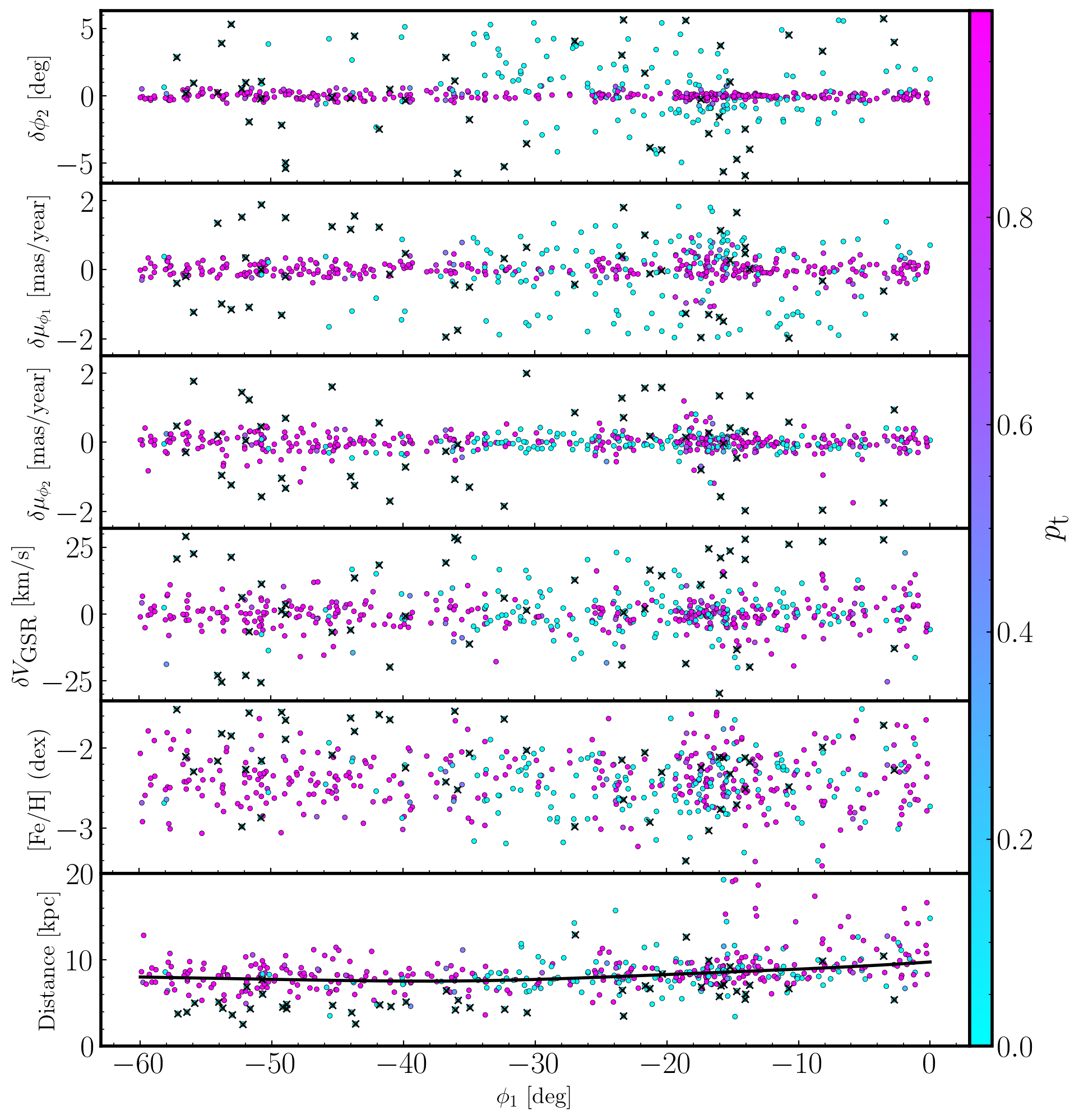}
    \caption{Parameters of DESI GD-1 member stars, plotted as a function of \phione\ between $-60^\circ<\phi_1<0^\circ$, the range used for the Gaussian mixture model. The top four panels (\dphitwo, \dpmone, \dpmtwo, and \dvgsr) are the four parameters used in the three component Gaussian mixture model used to characterize the thin stream, cocoon and background stars. The bottom two panels (\feh\ and distance) are not used in the mixture model. Points are colored by thin stream probability. Black crosses mark stars that have $p_{bg}>0.5$. The black line overlaid in the bottom panel is the distance track obtained in Section \ref{sec:dist-track}.}
    \label{fig:GMM-4panel}
\end{figure*}

\begin{figure*}
    \centering
    \includegraphics[width=\linewidth]{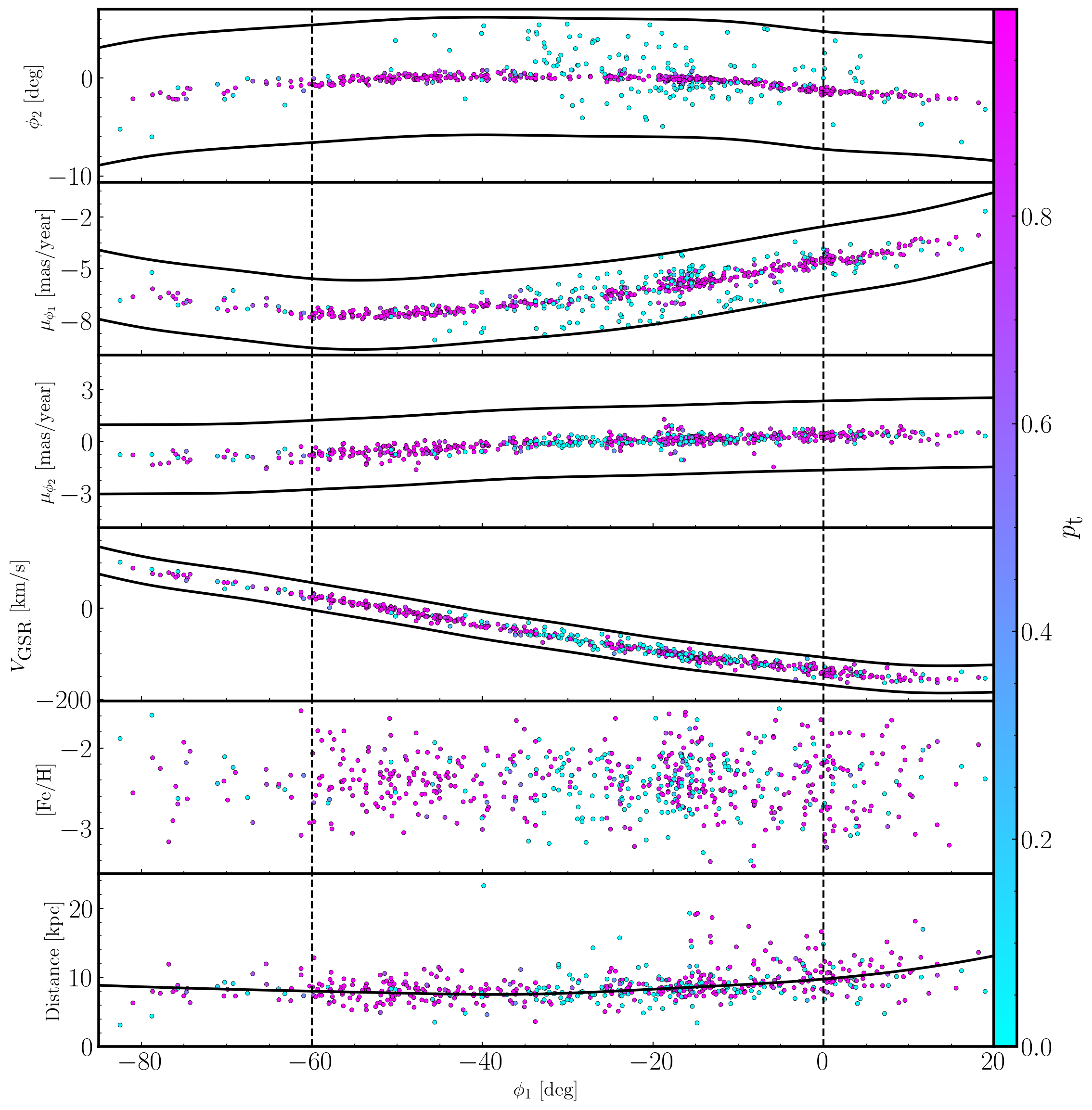}
    \caption{GD-1 member stars from DESI MWS coloured by stream probability with background stars ($p_{bg} > 0.5$) removed. Magenta points correspond to the highly probable thin stream members and cyan points correspond to the highly probable cocoon members. Parameters (\phitwo, \pmone, \pmtwo, \vgsr, \feh\ and distance) are plotted as a function of \phione\ along the entire length of the stream. In the top four panels, the black solid lines mark the selection around the stream track used to obtain the initial selection of member stars. Vertical dashed lines depict the region between $-60^\circ<\phi_1<0^\circ$ used in the stream-cocoon Gaussian mixture model. The black line overlaid in the bottom panel is the distance track obtained in Section \ref{sec:dist-track}.}
    \label{fig:p_stream_5panel}
\end{figure*}

Using the four-dimensional space ($\delta\phi_2$, $\delta\mu_{\phi_1}$, $\delta\mu_{\phi_2}$, $\delta V_{\rm GSR}$), with a three-component mixture (two Gaussians representing the thin stream and cocoon, and a uniform term for the foreground and background) mixture model, we calculate the probability that each star belongs to each of the three components using

\begin{equation}
p_{k,i} = \frac{f_k \, \prod_{j} P_{k,j}(x_{i,j})}{\displaystyle \sum_{m \in \{\mathrm{t,c,bg}\}} f_m \, \prod_{j} P_{m,j}(x_{i,j})},
\end{equation}

\noindent where $k \in \{\mathrm{t}, \mathrm{c}, \mathrm{bg}\}$ correspond to the thin stream, cocoon, and background components, respectively. Here $p_{k,i}$ is the posterior membership probability of star $i$ belonging to component $k$ and we refer to this as the membership probability of the star belonging to each of the three components. For each star, $i$,  $p_{t,i}+p_{c,i}+p_{bg,i}$ = 1. $f_k$ is the mixture fraction (prior weight) of that component. The quantity $x_{i,j}$ represents the observed variable $j$ for star $i$ (specifically $V_{\mathrm{GSR}}, \mu_{\phi_1}, \mu_{\phi_2}$, and $\phi_2$), and $P_{k,j}(x_{i,j})$ is the probability density of that variable under component $k$. Figure \ref{fig:GMM-4panel} displays the thin stream membership probabilities of the stars within $-60^\circ < \phi_1 < 0^\circ$ for the four parameters used in the Gaussian mixture model, along with \feh\ and distance.

We then extrapolate the results along $\phi_1$ to estimate membership probabilities for stars across the full extent of the stream ($-90^\circ < \phi_1 < 20^\circ$), assuming the model parameters remain unchanged over this extended range.\footnote{We caution that this assumption is generally invalid, particularly for the mixture fraction $f_k$. Membership probabilities in the extended $\phi_1$ range should therefore be interpreted with care. This extrapolation is used solely to provide a complete visualization of GD–1 membership probabilities and is not used in any subsequent analysis.} 
To obtain a high-purity catalog, we remove all stars with $p_{bg}>0.5$. With this cut, we obtain a final catalog of \comment{608} highly probable GD-1 member stars. We assign stars to the thin stream if they have $p_{t}>0.5$, and similarly, to the cocoon component if they have $p_{c}>0.5$. With these criteria, we find \comment{421} members of the thin stream, \comment{185} members of the cocoon, and \comment{73} background contaminants. As an alternative estimate, we sum the individual membership probabilities across all stars for each component, which provides a statistically expected number of members that accounts for stars with ambiguous classifications.
This yields \comment{407} thin stream members, \comment{194} cocoon members, and \comment{78} contaminants. The close agreement between these two approaches indicates that the model cleanly separates the three populations, with few stars having ambiguous membership assignments.
In Figure \ref{fig:p_stream_5panel} we show the \phitwo, \pmone, \pmtwo, \vgsr, \feh\ and distance of selected member stars (with $p_{bg}>0.5$ removed), color-coded by thin stream probability $p_t$. We clearly see a broad cocoon in \phitwo\ and \pmone\ between $-60^\circ<\phi_1<0^\circ$ with low $p_t$. All thin stream and cocoon membership probabilities are included in Table \ref{tab:gd1-members}.

\subsection{Metallicity of the thin stream and cocoon}\label{sec:GMM-feh}

\begin{figure}
    \centering
    
    \includegraphics[width=0.48\linewidth]{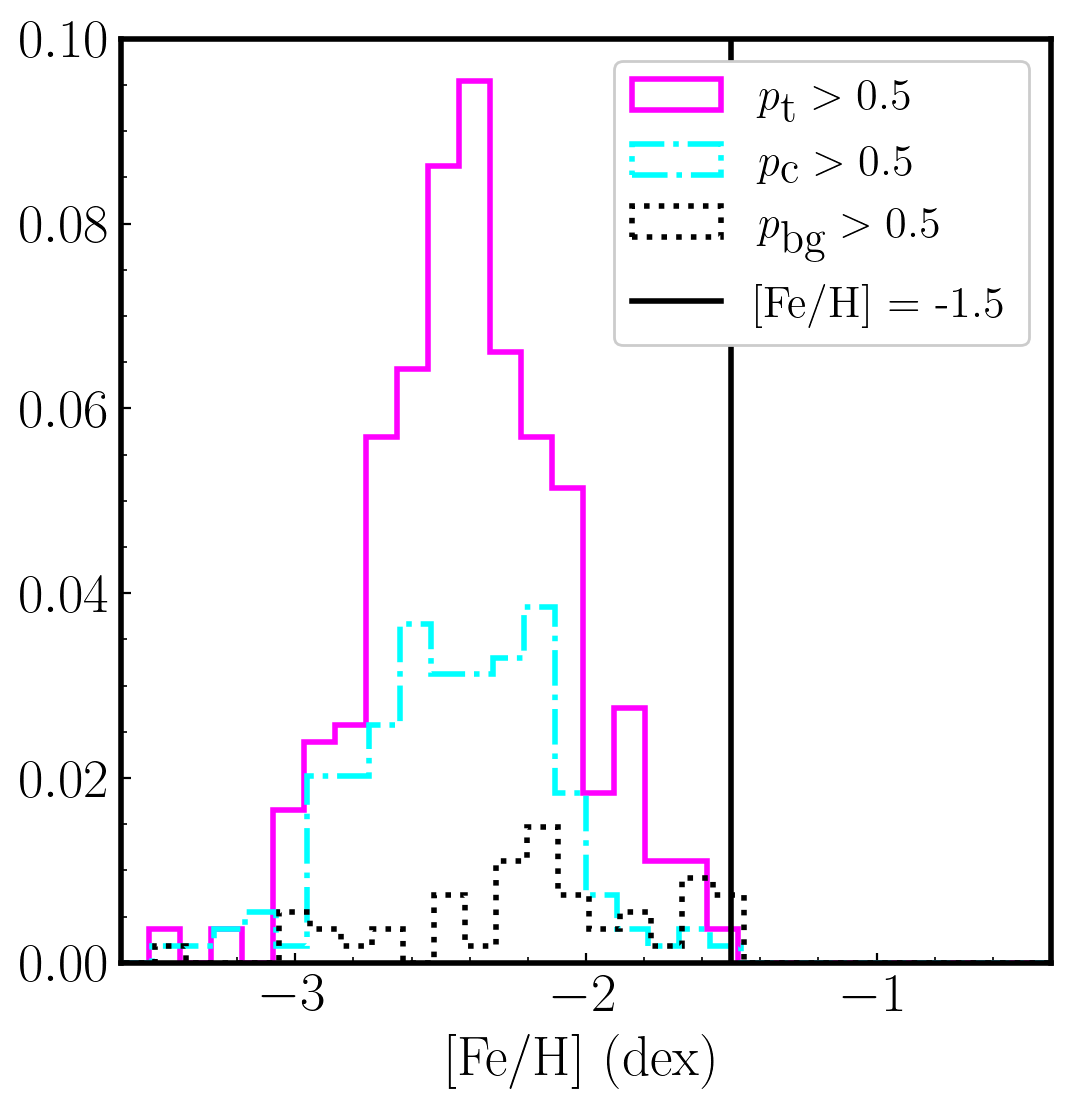}
    \includegraphics[width=0.48\linewidth]
    {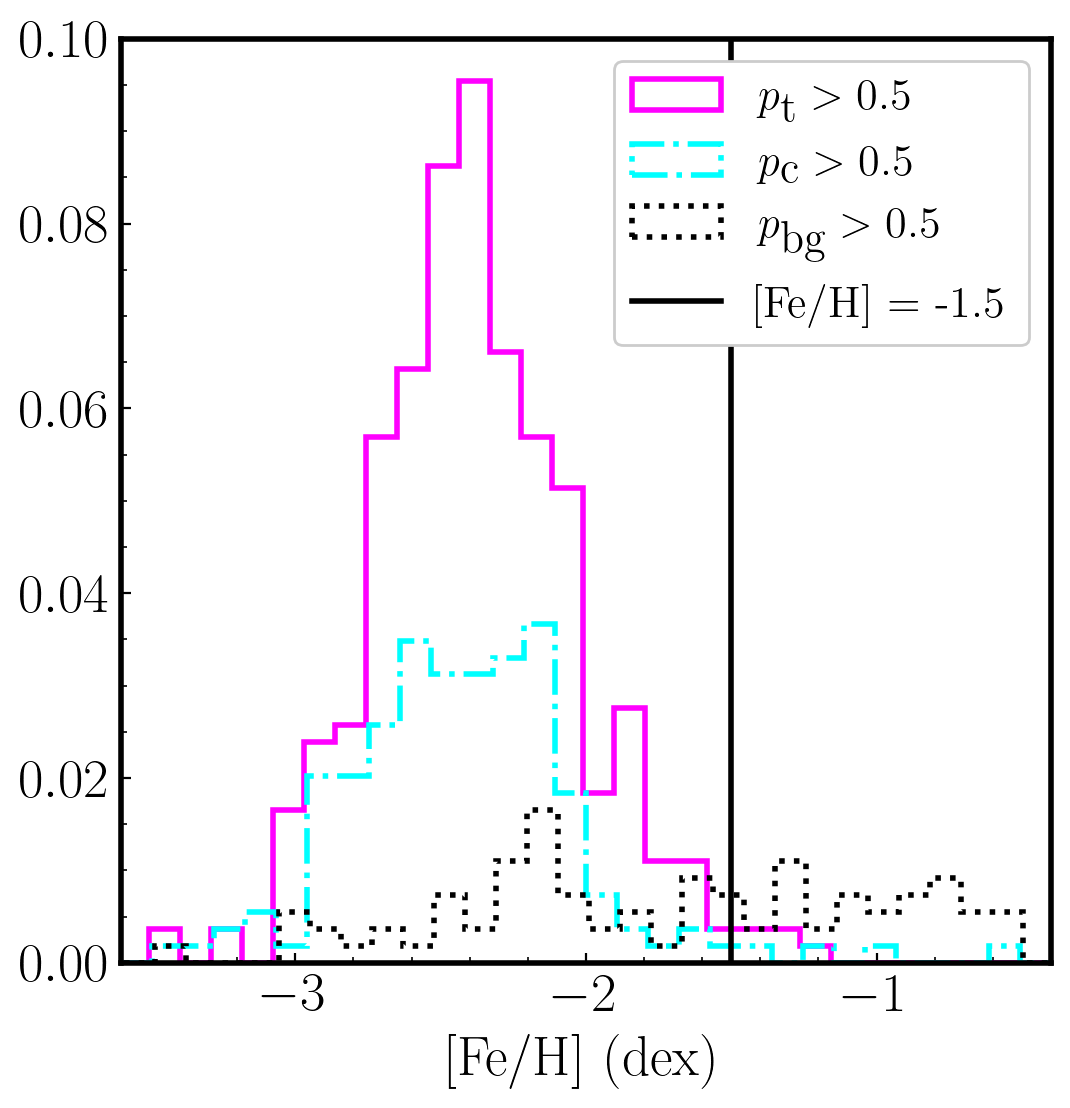}
    
    \caption{Metallicity distributions of GD-1 stars used in the Gaussian mixture model described in Section \ref{sec:GMM-description}. In each panel, the pink line indicates the high-probability thin stream members, the cyan dot-dashed line marks the highly-probable cocoon members and the dotted black line marks the highly-probable background contaminants. The solid black line marks the \feh~$=-1.5$ cut used in the DESI complete selection. The histograms are normalized by the total number of stars with \feh$<-1.5$, so that the integrated probability below this threshold equals one. \textit{Left panel}: Gaussian mixture model run on stars with an initial selection of [Fe/H] $<-1.5$. \textit{Right panel}: Gaussian mixture model run on stars with no initial selection in [Fe/H], showing that the model is able to isolate high metallicity stars as contaminants.}
    \label{fig:feh-dist}
\end{figure}

We then examine the metallicity distribution of the stream components to further characterize their distinct properties. 
The left panel of Figure \ref{fig:feh-dist} shows the metallicity distribution of the thin stream, cocoon and background populations obtained from our Gaussian mixture model with the cut of \feh\ $<-1.5$. To characterize the intrinsic metallicity of each population, we fit the member stars with a Gaussian distribution and use a Markov Chain Monte Carlo (MCMC) sampler with 50 walkers and 500 steps to obtain the posterior distribution of the mean and intrinsic dispersion, accounting for the individual \feh\ measurement uncertainties. For the highly-probable GD-1 member stars ($p_{bg} < 0.5$), we find an intrinsic mean metallicity of $\langle[\mathrm{Fe/H}]\rangle=-2.42\pm0.01$~dex and an intrinsic metallicity dispersion of $\sigma_{\mathrm{[Fe/H]}}=0.32\pm0.01$~dex. For the thin stream population ($p_t>0.5$), we find $\langle[\mathrm{Fe/H}]\rangle_t=-2.41\pm0.02$~dex and $\sigma_{\mathrm{[Fe/H], t}}=0.32\pm0.01$~dex, and for the cocoon ($p_c>0.5$), we find $\langle[\mathrm{Fe/H}]\rangle_c=-2.45\pm0.02$~dex and $\sigma_{\mathrm{[Fe/H], c}}=0.32\pm0.02$~dex. These values are consistent, within the uncertainties, with the metallicities reported by \citet{valluri_2025} based on the DESI EDR sample.

The measured metallicity dispersion of GD-1 is substantially larger than the typical dispersion of globular cluster streams (0.03--0.05~dex; \citealt{Li_2022}), and is more consistent with the dispersions observed in dwarf galaxies ($\gtrsim 0.25$~dex; \citealt{Li_2022}). This would favor a dwarf-galaxy-like progenitor for GD-1 over a classical globular cluster origin. However, we caution that the reported \feh\ uncertainties from the DESI pipeline are formal errors only and likely underestimate the true measurement uncertainty in the extremely metal-poor regime, where systematic effects in spectral modeling become significant. Since our method infers the intrinsic dispersion by deconvolving the measurement uncertainties, underestimated errors would lead to an overestimated intrinsic metallicity dispersion. Such systematics have been shown to artificially inflate the inferred intrinsic dispersion for chemically homogeneous systems: several well-studied, metal-poor globular clusters and streams in the DESI DR1 Stellar Catalogue (e.g., M\,92, M\,15, and NGC\,5053, all with \feh$<-2.0$~dex) exhibit metallicity dispersions substantially larger than their literature values (see Table 7 and Appendix D of \citealt{Koposov_2025}). A similar caveat has recently been discussed for the metal-poor C-19 stream analyzed with DESI \citep{mohammed_2026}. We therefore interpret our measured dispersion as an upper bound on the true intrinsic metallicity spread of GD-1, and while this value is consistent with a dwarf-galaxy-like progenitor, it alone is not conclusive.

To statistically test whether the metallicity distributions of the thin stream, cocoon, and background components originate from the same underlying distribution, we perform two-sample Kolmogorov-Smirnov tests. The result shows no statistical difference between the stream and cocoon components (p-value $\sim$0.31) which suggests that they share a similar underlying metallicity distribution. In contrast, both the stream and cocoon differ strongly from the background population, with p-values $<0.01$ which allows us to reject the null hypothesis in both cases. These results show that stream and cocoon likely originate from the same progenitor, while background stars represent a distinct population of contaminants.

To test the robustness of our component separation, we also apply our three-component kinematic model to a selection of stars without any prior cut in \feh. As shown in the right panel of Figure \ref{fig:feh-dist}, the model successfully isolates the majority of metal-rich background stars from the stream and cocoon components without any cut in \feh. 
The stream and cocoon remain relatively uncontaminated, with very few high-probability members at ${\rm \feh} > -1.5$, while the background component captures the majority of metal-rich stars.

\section{Discussion} \label{sec:discussion}

In this section, we discuss the results of the paper. First, in Section \ref{sec:comparison}, we compare our work with the previous studies of the GD-1 stream. Then, in Section \ref{sec:pm_disp}, we discuss the large observed dispersion in \pmone.

\subsection{Comparison with previous GD-1 studies}\label{sec:comparison}

This work dramatically increases the sample of spectroscopically confirmed GD-1 members. Using data from the DESI EDR, \citet{valluri_2025} identified 115 members of GD-1, but the EDR had minimal coverage of the GD-1 region, concentrated between $-30^\circ < \phi_1 < 0^\circ$ and within $3^\circ$ of the $\phi_2$ track. The DR2 data offers substantially more complete coverage along the entire stream. Prior to this work, the largest spectroscopic sample was presented by \citet{tavangar_2025}, which compiled 353 high-probability GD-1 members with radial velocity measurements from multiple surveys including DESI (EDR), SDSS, LAMOST, and targeted follow-up programs. However, this compilation combines data with heterogeneous selection functions, velocity precisions, and systematic uncertainties. Our DESI DR2 catalog provides \comment{608} spectroscopically confirmed members -- the largest \emph{homogeneous} sample of GD-1 to date, obtained from a single instrument with uniform target selection and consistent velocity precision. This represents a five-fold increase over the DESI EDR catalog and nearly doubles the total number of GD-1 members with spectroscopic measurements.

We cross match our catalog of likely GD-1 member stars with the catalogs from \cite{ibata_2024} and \cite{tavangar_2025} using the {\it Gaia} DR3 source ID of each star, as presented in Table \ref{tab:xmatch_summary}. Both of these catalogs identify high-probability GD-1 members primarily based on {\it Gaia} proper motions and parallaxes, without the benefit of radial velocity or metallicity information. The \cite{tavangar_2025} catalog contains 1689 high-probability members, of which 442 match stars in our high-probability catalog (391 thin stream, 49 cocoon). Of their 353 stars with radial velocity measurements, 163 overlap with our sample (127 thin stream, 35 cocoon). The \cite{ibata_2024} catalog contains 1468 GD-1 members, of which 436 match our high-probability catalog (354 thin stream, 81 cocoon).

In our spectroscopic sample, the ratio of thin stream to cocoon members is roughly 2:1. However, among the matched stars from the \cite{tavangar_2025} and \cite{ibata_2024} catalogs, the cocoon is significantly underrepresented: the stream-to-cocoon ratios are 8:1 and 4.4:1, respectively, far exceeding our measured 2:1. This deficit is not surprising, as proper-motion-based searches alone lack the sensitivity to efficiently identify cocoon members. The cocoon is a diffuse, kinematically hotter structure; radial velocities and metallicities -- available in our DESI sample but absent from those photometric and astrometric catalogs -- are critical for cleanly separating cocoon stars from contaminants. This implies that the true GD-1 population, particularly the cocoon component, is likely substantially larger than reported in those catalogs, and that spectroscopic surveys such as DESI are essential for obtaining a complete census of the stream and its surrounding structure.

\begin{table*}
\centering
\caption{Cross-match summary between our catalog, Tavangar \& Price-Whelan (2025), and Ibata et al. (2024).}
\label{tab:xmatch_summary}
\begin{tabular}{lcccc}
\toprule
Catalog & Total stars & \multicolumn{3}{c}{Matched to our catalog} \\
\cmidrule(lr){3-5}
 &  & $p_t+p_c>0.5$ & $p_t>0.5$ & $p_c>0.5$ \\
\midrule
\midrule
This work & 679 & 608 & 421 & 185 \\
Tavangar \& Price-Whelan (2025) & 1689 & 442 & 391 & 49 \\
Tavangar \& Price-Whelan (2025) with RV & 353 & 163 & 127 & 35 \\
Ibata et al. (2024) GD-1 & 1468 & 436 & 357 & 77 \\
\bottomrule
\end{tabular}
\end{table*}

\citet{valluri_2025} also perform a bivariate Gaussian mixture model analysis using $\delta\phi_2$ and \dvgsr\ to characterize the thin stream and cocoon components of GD-1. With both the DESI-EDR and DR2 observations, a similar fraction of stars is found to be part of the thin stream ($0.60\pm0.02$ and $0.61\pm0.05$ for the DR2 and EDR data, respectively), although \citet{valluri_2025} do not account for background contamination nor proper motion measurements in their model. For both data sets, a larger width and velocity dispersion are observed for the cocoon compared to the thin stream. The values of the thin stream and cocoon width are both larger for the DR2 data than the EDR data, likely because the EDR data used a narrower \phitwo\ cut of $3^\circ$ (compared with $6^\circ$ in this work) and had sparser, less uniform coverage along the stream (see Figure~\ref{fig:desi-releases}). The velocity dispersions of both the thin stream and cocoon are smaller in our DR2 analysis compared with the EDR results. This difference can likely be attributed to the larger number of member stars obtained in this work as well as the inclusion of the proper motions in the Gaussian mixture model.

\begin{figure*}
    \centering
    \includegraphics[width=\linewidth]{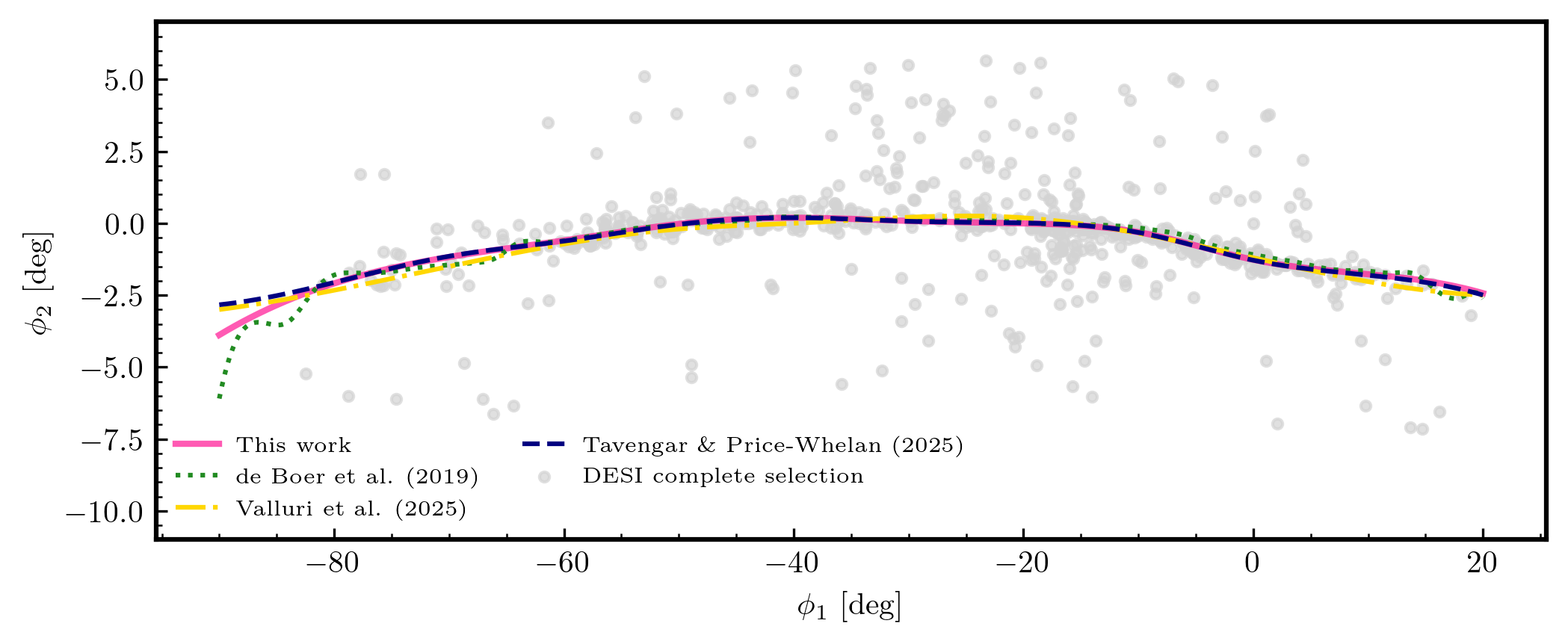}
    \caption{Comparison of the $\phi_2$  stream track derived in this work (pink) with the tracks from \citet{valluri_2025} (yellow), \citet{deboer_2020} (green) and \citet{tavangar_2025} (navy). Grey points mark our DESI complete selection of GD-1 member stars.}
    \label{fig:phi2-tracks}
\end{figure*}

Several previous works have derived stream tracks for GD-1 \citep{tavangar_2025, valluri_2025, deBoer_2018}. We compare our $\phi_2$ track with these studies in Figure~\ref{fig:phi2-tracks}. Our track closely matches that of \citet{tavangar_2025} over most of the stream, with small deviations at $\phi_1 < -80^\circ$ where the member sample is sparse. 

\begin{figure*}
    \centering
    \includegraphics[width=\linewidth]{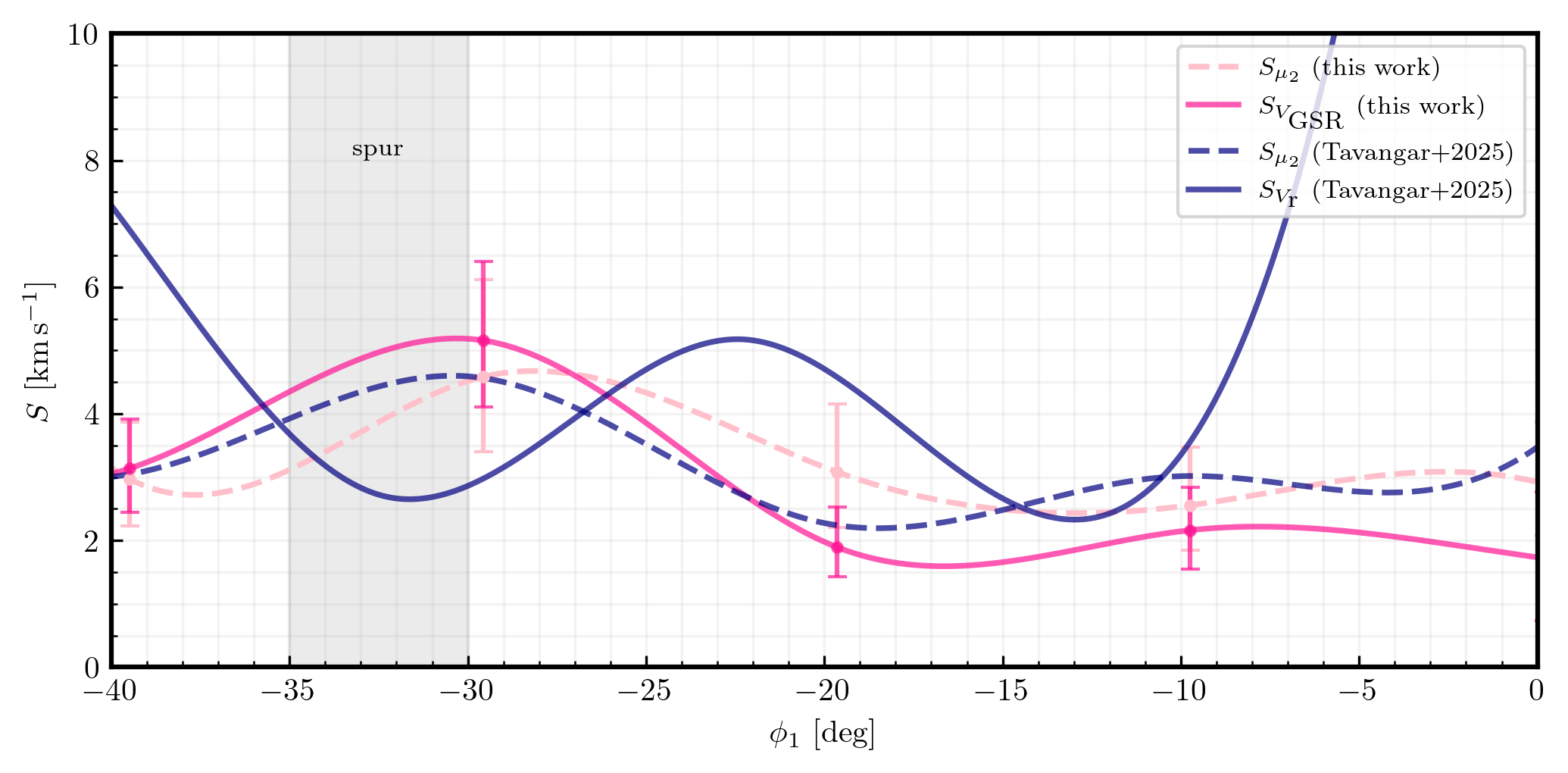}
    \caption{Comparison of velocity dispersions along the stream between this work (pink) and \citet{tavangar_2025} (navy). Solid lines show the line-of-sight velocity dispersion ($S_{V_{\rm GSR}}$) and dashed lines show the transverse velocity dispersion perpendicular to the stream ($S_{\mu_{\phi_2}}$), converted to \kms\ using our distance track (Section~\ref{sec:dist-track}). Pink points with error bars indicate the spline node values from our \texttt{Stan} model fits (see Section~\ref{sec:pm_sel}). The shaded grey region marks the location of the spur identified by \citet{Price-Whelan_2018}. Both $S_{V_{\rm GSR}}$ and $S_{\mu_{\phi_2}}$ from this work show peaks near these features, while the \citet{tavangar_2025} $S_{\mu_{\phi_2}}$ track shows a similar trend. Their $S_{V_{\rm GSR}}$ track differs from ours, likely due to heterogeneous radial velocity sources in their compilation.}

    \label{fig:dispersion}
\end{figure*}

In Figure~\ref{fig:dispersion}, we examine how the velocity dispersion varies along the stream. Both the line-of-sight velocity dispersion ($S_{V_{\rm GSR}}$) and the transverse velocity dispersion perpendicular to the stream ($S_{\mu_{\phi_2}}$, converted to \kms\ using our distance track) probe motion perpendicular to the stream's orbit, so we expect them to trace similar dynamical heating. Indeed, our measurements of these two quantities show remarkably similar trends: both exhibit one peak at a similar location along the stream, near $\phi_1 \approx -30^\circ$. These peaks coincide with the previously identified spur \citep{Price-Whelan_2018}, suggesting a possible connection between spatial perturbations and localized kinematic heating.

To assess whether this structure is robust, we compare our $S_{\mu_{\phi_2}}$ measurements with those from \citet{tavangar_2025}. Since both analyses use \textit{Gaia} proper motions, this provides an independent check on our methodology. The two measurements are consistent within $1\sigma$, and show similar peak locations. In contrast, the $S_{V_{\rm GSR}}$ track from \citet{tavangar_2025} differs significantly from ours. This discrepancy likely arises because \citet{tavangar_2025} compiled radial velocities from multiple surveys with heterogeneous systematics and velocity precisions, whereas our DESI sample provides homogeneous measurements with consistent uncertainties.

We caution that the amplitude of the dispersion variations is comparable to the measurement uncertainties, so the peak structure could be consistent with a constant dispersion at $2\sigma$. Additionally, the precise locations of the peak depend on the placement of spline nodes in our model, and can shift slightly with different node choices. Nevertheless, the agreement between the line-of-sight and transverse velocity dispersions, combined with the spatial coincidence with the spur feature, hints at localized kinematic substructure that may be connected to past perturbations. Confirming these features will require larger spectroscopic samples from future DESI data releases and better proper motion measurements from future Gaia releases.

\begin{figure}[!htpb]
    \centering
    \includegraphics[width=\linewidth]{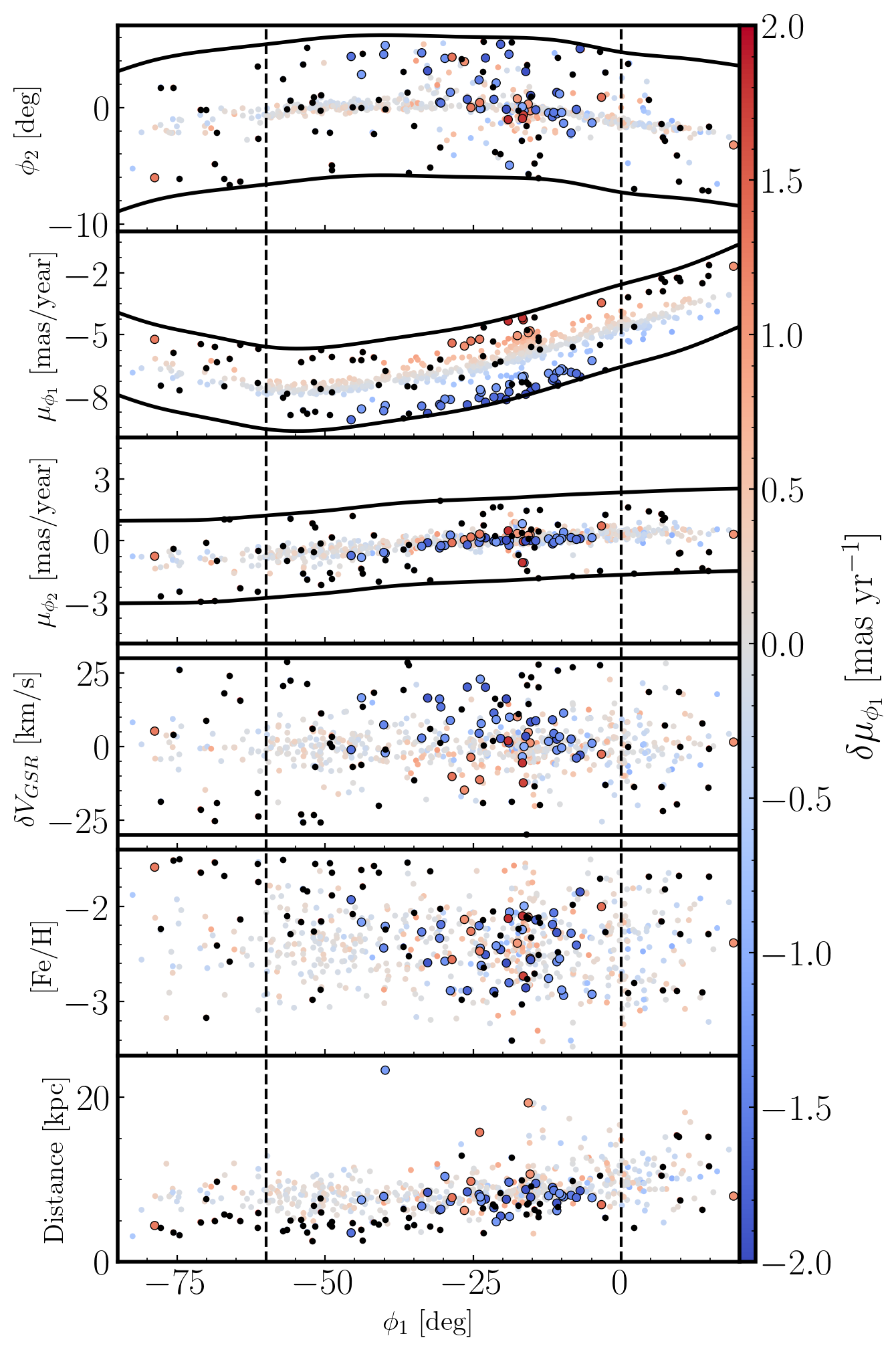}
    \caption{Six-panel summary of GD-1 stream stars selected by DESI, showing stellar properties as a function of \phione, colored by the proper motion offset \dpmone. Stars with $p_{\rm bg} > 0.5$ are marked with black points. Black-edged markers outline stars with $|$\dpmone$| > 1$~\masyr, which contribute most to the large cocoon dispersion. These high-offset stars occupy the same locus as the thin stream in \pmtwo, \feh, and distance, indicating they are genuine GD-1 members. In the top four panels, solid black lines show the selection boundaries used in the initial member selection. Vertical dotted lines mark the region $-60^\circ < \phi_1 < 0^\circ$ used in the Gaussian mixture model analysis.}
    \label{fig:6panel_pmphi1}
\end{figure}

\begin{figure}[!htpb]
    \centering
    \includegraphics[width=\linewidth]{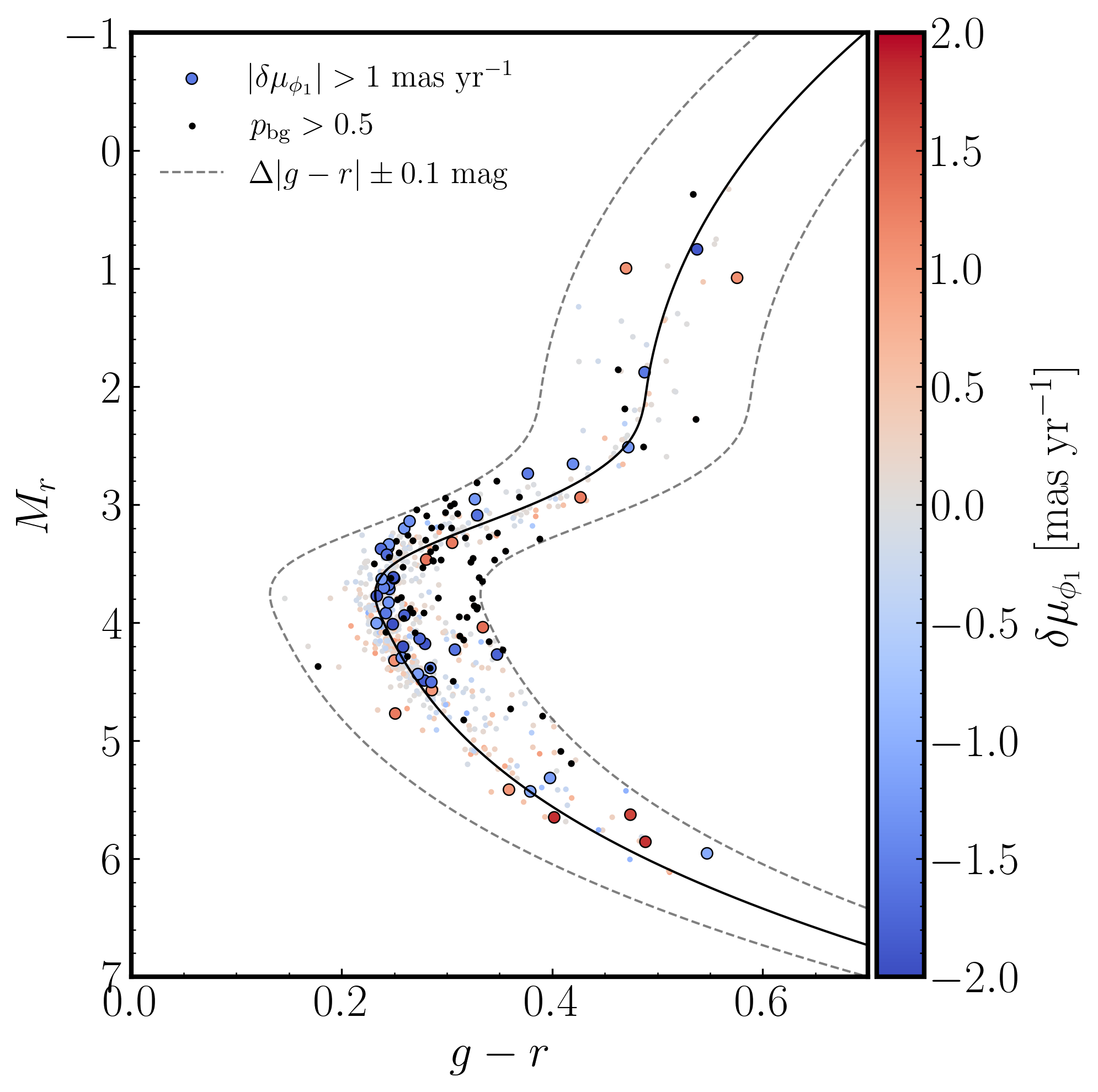}
    \caption{Color-magnitude diagram of DESI GD-1 stars, colored by the proper motion offset \dpmone. Stars with $p_{\rm bg} > 0.5$ are marked with black points. Black-edged markers outline stars with $|$\dpmone$| > 1$~\masyr. These high-offset stars follow the same isochrone as the rest of the stream, confirming their membership. Stars with negative \dpmone\ (below the \pmone\ track) preferentially lie above the isochrone (closer), while stars with positive \dpmone\ lie below (more distant), suggesting that the large \pmone\ dispersion reflects a line-of-sight distance spread. The isochrone and dashed selection boundaries ($\Delta|g-r| = \pm0.1$~mag) are the same as in Figure~\ref{fig:cmd}.}
    \label{fig:cmd_dpmphi1}
\end{figure}

\begin{figure*}
    \centering
    \includegraphics[width=\linewidth]{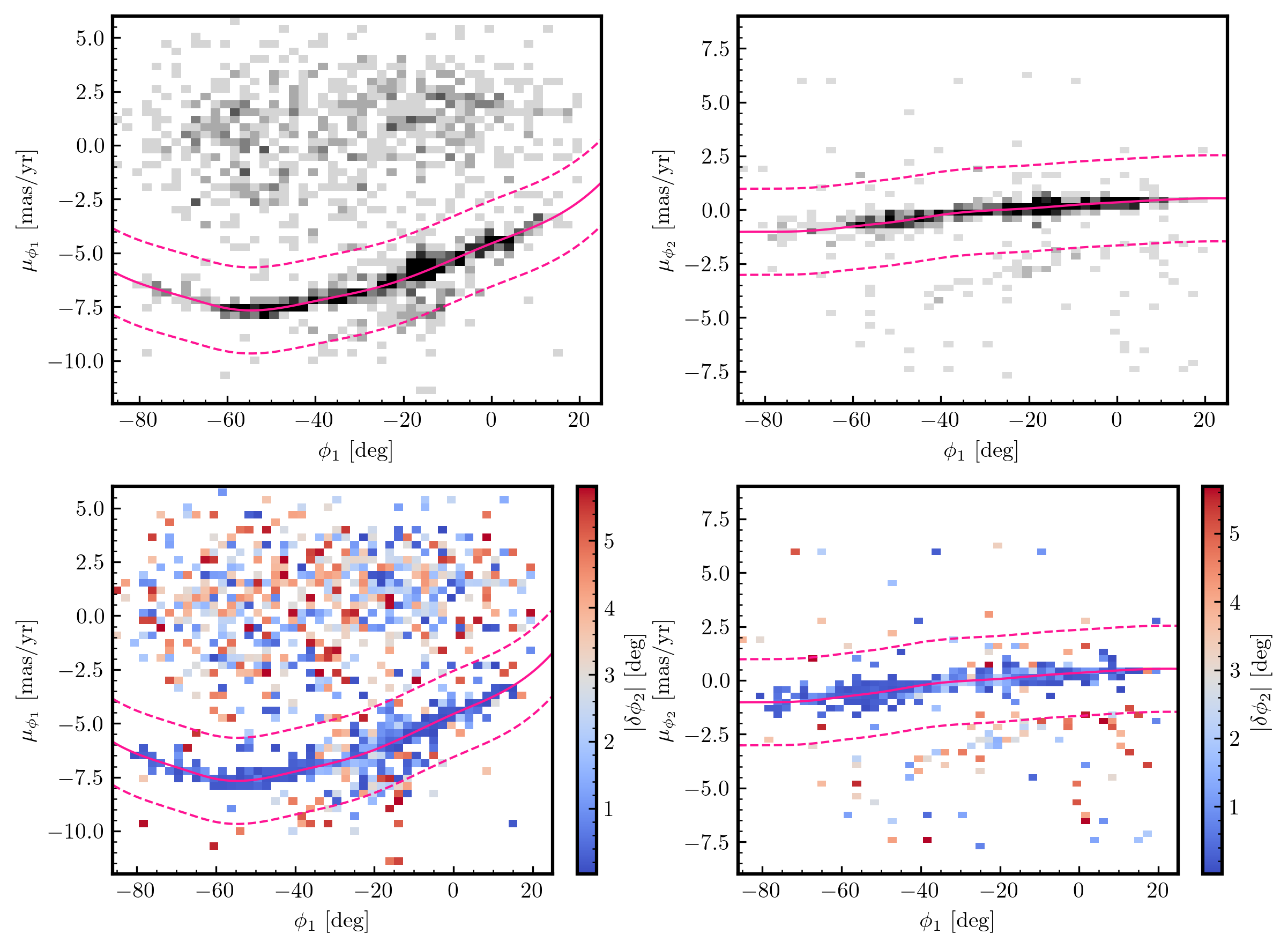}
    \caption{Proper motion distributions of DESI GD-1 stars with the complete selection in CMD, \phitwo, \vgsr, and \feh~$<-2$, but without the $|$\dpmone$| < 2$~\masyr\ restriction. Left panels show \pmone\ vs.\ \phione; right panels show \pmtwo\ vs.\ \phione. Top panels display the stellar surface density; bottom panels are colored by $|\delta\phi_2|$. Solid pink lines show the stream tracks and dashed lines indicate the $\pm2$~\masyr\ selection boundary used in Section~\ref{sec:mem_sel_summary}. In the left panels, a ``cloud'' of stars extends below the \pmone\ track between $-50^\circ \lesssim \phi_1 \lesssim 0^\circ$, beyond the selection boundary. This asymmetric feature is not present in \pmtwo\ (right panels). The bottom-left panel shows that stars with larger \pmone\ offsets also have larger $|\delta\phi_2|$, indicating a connection between kinematic and spatial structure.}
    \label{fig:pm_2x2}
\end{figure*}

\subsection{The \dpmone\ cocoon dispersion}\label{sec:pm_disp}

As shown in Section~\ref{sec:GMM}, we measure a large \dpmone\ dispersion of \comment{$41.36\pm4.98$~\kms} for the cocoon component of the GD-1 stream. Such a value is not typical for a dynamically cold stellar stream in a CDM halo, which usually has velocity dispersions of only a few \kms\ \citep[e.g.][]{Li_2019, Malhan_Ibata_2019, Malhan_2022}. We therefore must determine whether this measurement reflects a real property of GD-1 or is an artifact of contamination.

Figures~\ref{fig:6panel_pmphi1} and \ref{fig:cmd_dpmphi1} provide evidence that the stars contributing to this large dispersion are genuine GD-1 members. In Figure~\ref{fig:6panel_pmphi1}, we highlight stars with large deviations from the \pmone\ track ($|$\dpmone$| > 1$~\masyr), which contribute most to the measured dispersion. These stars occupy the same locus as the thin stream in \pmtwo, \feh, and distance, and show no distinguishing features that would indicate contamination. Figure~\ref{fig:cmd_dpmphi1} reinforces this conclusion: stars with high $|$\dpmone$|$ follow the same isochrone as the rest of the stream population along the main sequence and subgiant branch. If these stars were contaminants, they would not trace the GD-1 sequence so precisely. This interpretation is further supported by our metallicity analysis (Section~\ref{sec:GMM-feh}), which shows that the cocoon shares the same metallicity distribution as the thin stream. Apart from their large \pmone\ offset, there is no reason to distinguish these stars from genuine GD-1 members.

Since the mixture model analysis in Section~\ref{sec:GMM} is restricted to member candidates satisfying $|$\dpmone$| < 2$~\masyr, the measured cocoon dispersion is limited by this selection boundary. To investigate whether additional members exist beyond this cut, we show the proper motion distribution of GD-1 stars in Figure~\ref{fig:pm_2x2}, applying the complete selection in \phitwo, \vgsr, CMD, and \feh, but extending the \pmone\ range. The left panels reveal a broad feature in \pmone, particularly between $-50^\circ \lesssim \phi_1 \lesssim 0^\circ$. Notably, this broadening is asymmetric: a ``cloud'' of stars lies systematically below the \pmone\ track (at smaller \pmone\ values, roughly $-10~\mathrm{mas~yr}^{-1} < \mu_{\phi_1} < -7~\mathrm{mas~yr}^{-1}$), extending beyond the \dpmone~$= -2$~\masyr\ selection boundary. In contrast, the right panels show no such offset or broadening in \pmtwo, consistent with the much smaller cocoon dispersion measured in that direction. The bottom panels of Figure~\ref{fig:pm_2x2} show that stars with larger $|$\dpmone$|$ offsets also tend to have larger $|\delta\phi_2|$, indicating a connection between the spatial width and kinematic structure of the cocoon.

We defer a thorough study of this large \pmone\ dispersion, along with its origin and interpretation, to a forthcoming DESI publication, but briefly outline a likely explanation below. The quantities \pmone\ and \pmtwo\ here are proper motions corrected for solar reflex motion, which requires an assumed distance to each star. In this work, we adopt the mean stream distance at each $\phi_1$ (i.e.\ the distance track we build in Section~\ref{sec:dist-track}). However, if the cocoon has a significant spread in line-of-sight distance, stars at different distances will have their solar reflex corrections systematically biased: stars closer than assumed will appear to have smaller \pmone\ (negative \dpmone), while more distant stars will appear to have larger \pmone\ (positive \dpmone). This effect would manifest as an apparent \pmone\ dispersion even in the absence of true tangential velocity spread.

To estimate whether line-of-sight distance spread could account for the large \pmone\ dispersion, we recompute the solar-reflex-corrected proper motions after perturbing the observed distance track, thereby measuring the sensitivity of \pmone\ to the assumed distance. We find that a distance spread
on the order of $\sim$15\% is needed to produce the observed \pmone\ broadening through the reflex correction alone. For comparison, the on-sky cocoon width of $\sim2.2^\circ$ corresponds to only $\sim0.3$~kpc at the stream distance of 8~kpc, whereas a 15\% distance spread implies a line-of-sight extent of $\sim1.2$~kpc --- roughly four times larger. This would indicate that the cocoon is substantially more extended along the line of sight than on the plane of the sky.

This interpretation is also supported by Figure~\ref{fig:cmd_dpmphi1}, which shows a clear correlation between \dpmone\ and position on the color-magnitude diagram. At the main-sequence turn-off, stars with negative \dpmone\ (below the \pmone\ track) preferentially lie above the isochrone, indicating they are closer than the assumed stream distance, while stars with positive \dpmone\ lie below the isochrone, indicating greater distances. The magnitude offset between these two populations at the turnoff is $\sim$0.3~mag, consistent with a distance difference of $\sim$10--15\%, in broad agreement with the estimate above. The asymmetry of the ``cloud''---predominantly at negative \dpmone---is consistent with the cocoon being systematically closer than the thin stream along the line of sight.

We note that the solar reflex correction also affects \pmtwo, with a sensitivity to distance that is approximately half that of \pmone\
at the sky position of GD-1. A 15\% distance spread would therefore predict an additional \pmtwo\ dispersion of $\sim$15--20~km~s$^{-1}$, which is not evident in our measured cocoon \pmtwo\ dispersion of $2.81\pm 0.58$~km~s$^{-1}$. While the smaller amplitude of the distance-induced effect in \pmtwo\ makes it inherently harder to detect than in \pmone, the expected broadening still exceeds the typical proper motion uncertainties of the GD-1 members ($\sim$0.15~mas~yr$^{-1}$ or 6~\kms) and should in principle be measurable. This suggests that the large \pmone\ dispersion may not be entirely attributable to distance spread, but rather a combination of line-of-sight distance spread and some degree of intrinsic velocity dispersion. Alternatively, it may indicate a correlation between transverse velocity and distance among the cocoon stars that acts to suppress the apparent dispersion in \pmtwo\ but not in \pmone.

We therefore suggest that the large \pmone\ dispersion is at least partially a consequence of line-of-sight distance spread in the cocoon. If confirmed, this would imply that the cocoon's spatial extent is three-dimensional, with significant depth along the line of sight. A detailed analysis disentangling the contributions of distance spread and intrinsic velocity dispersion, along with the possible origin of the ``cloud,'' will be presented in a forthcoming paper with a larger GD-1 sample from DESI (DESI Collaboration, in prep).

\section{Summary} \label{sec:summary}
We have assembled the largest homogeneous spectroscopic sample of GD-1 stars to date using DESI DR2, doubling the number of spectroscopically confirmed members compared to previous compilations that combined data from multiple instruments with inhomogeneous systematics. We detect clear evidence of a narrow stream component and a broader population (cocoon) with 30\% of stars in the broad component. The large sample of member stars has enabled tighter constraints on the width and velocity dispersion of both stream components. When combined with simulations, these measurements will be able to provide constraints on the distribution of dark matter subhalos in the Milky Way. 

We summarize the results of our paper as follows: 

\begin{enumerate}
    \item We present a spectroscopic analysis of the GD-1 stellar stream using the first three years of DESI observations (i.e., DESI DR2). 
    
    \item 
   
    We derive updated stream tracks in sky position (\phitwo), proper motion (\pmone, \pmtwo), and radial velocity (\vgsr) using a mixture model that allows the mean and dispersion to vary along $\phi_1$ (Table~\ref{tab:Stan_tracks}). Combined with an improved distance track and CMD selection, we identify \comment{679} candidate members (Section~\ref{sec:mem_sel_summary}).
    
    \item The thin stream, cocoon, and background contaminant components are further modeled with a mixture model on the candidate members. We confirm the existence of a broad (\comment{$2.18^\circ\pm0.17^\circ$} $\approx743$~pc), kinematically hot (\comment{$6.13\pm0.75$})~\kms\ cocoon component surrounding a thin (\comment{$0.23^\circ\pm0.01^\circ$} $\approx78$~pc), cold (\comment{$2.49\pm0.28$})~\kms\ stream (see Table \ref{tab:gmm_comparison} in Section \ref{sec:GMM}). 

    \item 
    We detect a large proper motion dispersion along the stream ($\comment{41.36\pm4.98}$~\kms, or $1.13\pm0.14$~\masyr.) in the cocoon component (see Section \ref{sec:pm_disp}). We show that this is likely a consequence of line-of-sight distance spread rather than tangential velocity heating. The details of this interpretation and the associated ``cloud'' of stars offset from the \pmone\ track will be explored in a forthcoming paper.
    
    \item Our final catalog of highly probable GD-1 member stars contains \comment{608} stars, including \comment{421} stars in the thin stream and \comment{185} stars in the cocoon component, providing the largest catalog of spectroscopically confirmed members of GD-1 to date. The final table of the member stars with the membership probability is included in Table \ref{tab:gd1-members} in Appendix \ref{sec:appendix-desi-table}.
\end{enumerate}

Our measured cocoon velocity dispersion of \comment{$6.13\pm0.75$}~\kms\ is in excellent agreement with the $6.2\pm1.7$~\kms\ predicted by \citet{carlberg_2025} for a stream heated by a cold dark matter subhalo 
population over $\sim11$~Gyr. This consistency supports the interpretation that the cocoon represents dynamically heated stream material, though distinguishing between heating 
mechanisms requires detailed comparison with simulations (Carlberg et al., in prep).

We expect future DESI data releases to continue to increase the number of GD-1 members. With DR2, the average DESI completeness across the GD-1 region is 44\% (see Figure \ref{fig:completeness}). We therefore expect the number of GD-1 member stars to increase by a factor of $\sim2$ when the DESI survey is complete. Expanding the analysis to include these new observations will further improve the characterization of the multiple components of the GD-1 stream, and improve our understanding of the origin of these components.

\begin{center}
    \textbf{Data Availability}
\end{center}
This work is supplemented by a preliminary Zenodo repository (DOI: \href{https://doi.org/10.5281/zenodo.19258456}{10.5281/zenodo.19258456}), from which the data behind all figures and tables presented in this paper are included in machine readable format. Results are subject to change after the referee report.

\begin{center}
    \textbf{Acknowledgments}
\end{center}

E. J. acknowledges the support of the Natural Sciences and Engineering Research Council of Canada (NSERC) through a Canada Graduate Scholarship - Doctoral (CGS D) award. E. J. and T.S.L. acknowledge financial support from Natural Sciences and Engineering Research Council of Canada (NSERC) through grant RGPIN-2022-04794.
S.K. acknowledges support from Science \& Technology Facilities Council (STFC) (grant ST/Y001001/1).
MV is supported  by NASA-ATP grant 80NSSC24K0938  and NSF grant 2510879.
LBeS acknowledges support from CNPq (Brazil) through a research
productivity fellowship, grant no. [304873/2025-0]. C.A.P. is thankful for funding from the Spanish government through grants AYA2014-56359-P, AYA2017-86389-P and PID2020-117493GB-100.

This material is based upon work supported by the U.S. Department of Energy (DOE), Office of Science, Office of High-Energy Physics, under Contract No. DE–AC02–05CH11231, and by the National Energy Research Scientific Computing Center, a DOE Office of Science User Facility under the same contract. Additional support for DESI was provided by the U.S. National Science Foundation (NSF), Division of Astronomical Sciences under Contract No. AST-0950945 to the NSF’s National Optical-Infrared Astronomy Research Laboratory; the Science and Technology Facilities Council of the United Kingdom; the Gordon and Betty Moore Foundation; the Heising-Simons Foundation; the French Alternative Energies and Atomic Energy Commission (CEA); the National Council of Humanities, Science and Technology of Mexico (CONAHCYT); the Ministry of Science, Innovation and Universities of Spain (MICIU/AEI/10.13039/501100011033), and by the DESI Member Institutions: \url{https://www.desi.lbl.gov/collaborating-institutions}. Any opinions, findings, and conclusions or recommendations expressed in this material are those of the author(s) and do not necessarily reflect the views of the U. S. National Science Foundation, the U. S. Department of Energy, or any of the listed funding agencies.

The authors are honored to be permitted to conduct scientific research on I'oligam Du'ag (Kitt Peak), a mountain with particular significance to the Tohono O’odham Nation.

This research has made use of the SIMBAD database, operated at CDS, Strasbourg, France.
This research has made use of NASA’s Astrophysics Data System Bibliographic Services.

This paper made use of the Whole Sky Database (wsdb) created by Sergey Koposov and maintained at the Institute of Astronomy, Cambridge by Sergey Koposov, Vasily Belokurov and Wyn Evans with financial support from the Science \& Technology Facilities Council (STFC) and the European Research Council (ERC).

This work has made use of data from the European Space Agency (ESA) mission
{\it Gaia} (\url{https://www.cosmos.esa.int/gaia}), processed by the {\it Gaia}
Data Processing and Analysis Consortium (DPAC,
\url{https://www.cosmos.esa.int/web/gaia/dpac/consortium}). Funding for the DPAC
has been provided by national institutions, in particular the institutions
participating in the {\it Gaia} Multilateral Agreement.

For the purpose of open access, the author has applied a Creative
Commons Attribution (CC BY) licence to any Author Accepted
Manuscript version arising from this submission.

The DESI Legacy Imaging Surveys consist of three individual and complementary projects: the Dark Energy Camera Legacy Survey (DECaLS), the Beijing-Arizona Sky Survey (BASS), and the Mayall z-band Legacy Survey (MzLS). DECaLS, BASS and MzLS together include data obtained, respectively, at the Blanco telescope, Cerro Tololo Inter-American Observatory, NSF’s NOIRLab; the Bok telescope, Steward Observatory, University of Arizona; and the Mayall telescope, Kitt Peak National Observatory, NOIRLab. NOIRLab is operated by the Association of Universities for Research in Astronomy (AURA) under a cooperative agreement with the National Science Foundation. Pipeline processing and analyses of the data were supported by NOIRLab and the Lawrence Berkeley National Laboratory (LBNL). Legacy Surveys also uses data products from the Near-Earth Object Wide-field Infrared Survey Explorer (NEOWISE), a project of the Jet Propulsion Laboratory/California Institute of Technology, funded by the National Aeronautics and Space Administration. Legacy Surveys was supported by: the Director, Office of Science, Office of High Energy Physics of the U.S. Department of Energy; the National Energy Research Scientific Computing Center, a DOE Office of Science User Facility; the U.S. National Science Foundation, Division of Astronomical Sciences; the National Astronomical Observatories of China, the Chinese Academy of Sciences and the Chinese National Natural Science Foundation. LBNL is managed by the Regents of the University of California under contract to the U.S. Department of Energy. The complete acknowledgments can be found at https://www.legacysurvey.org/acknowledgment/.

\appendix

\section{Table of DESI DR2 Measurements in the GD-1 Area}\label{sec:appendix-desi-table}
Table \ref{tab:gd1-members} presents the \comment{679} stars in the ``DESI complete selection" sample defined in Section~\ref{sec:mem_sel_summary}. For each star, we list DESI DR2 measurements of \vgsr\ and \feh\ with their 1$\sigma$ uncertainties, along with the {\it Gaia} DR3 \texttt{SOURCE\_ID}, position, and membership probabilities from the Gaussian mixture model described in Section~\ref{sec:GMM-description}.
In Table \ref{tab:bhb-members} we also provide a list of the 7 BHB stars identified in Section \ref{sec:cmd}.

\begin{table*}[h]
    \centering
    \begin{tabular}{lrrrrrrrrrr}
\toprule
\textit{Gaia} Source ID & R.A. & Dec & $\phi_1$ & $\phi_2$ & $V_{\mathrm{GSR}}$ & $V_{\mathrm{err}}$ & [Fe/H] & [Fe/H]$_\mathrm{err}$ & $p_\mathrm{thin}$ & $p_\mathrm{cocoon}$ \\
 & $\mathrm{[deg]}$ & $\mathrm{[deg]}$ & $\mathrm{[deg]}$ & $\mathrm{[deg]}$ & $\mathrm{[km~s^{-1}]}$ & $\mathrm{[km~s^{-1}]}$ & $\mathrm{[dex]}$ & $\mathrm{[dex]}$ &  &  \\
\midrule
793716935624289536 & $147.24$ & $33.61$ & $-42.87$ & $-0.22$ & $-24.08$ & $1.27$ & $-2.16$ & $0.03$ & $0.98$ & $0.02$ \\
802157439717465472 & $151.25$ & $38.00$ & $-37.41$ & $-0.09$ & $-43.06$ & $2.23$ & $-2.43$ & $0.06$ & $1.00$ & $0.00$ \\
794522430972601600 & $147.03$ & $34.30$ & $-42.42$ & $0.33$ & $-26.65$ & $3.90$ & $-2.23$ & $0.08$ & $1.00$ & $0.00$ \\
\bottomrule
\end{tabular}

    \caption{DESI-DR2 observations of the GD-1 region (first 3 rows of table). The full machine-readable table with all rows and additional columns is available on Zenodo (see Data Availability).}
    \label{tab:gd1-members}
\end{table*}

\begin{table*}[h]
    \centering
    \begin{tabular}{lrrrrrrrr}
\toprule
\textit{Gaia} Source ID & R.A. & Dec. & $V_{\mathrm{GSR}}$ & $V_{\mathrm{err}}$ & [Fe/H] & [Fe/H]$_\mathrm{err}$ & $g$ & $r$ \\
 & $\mathrm{[deg]}$ & $\mathrm{[deg]}$ & $\mathrm{[km~s^{-1}]}$ & $\mathrm{[km~s^{-1}]}$ & $\mathrm{[dex]}$ & $\mathrm{[dex]}$ & $\mathrm{[mag]}$ & $\mathrm{[mag]}$ \\
\midrule
831434960462203904 & $163.13$ & $47.51$ & $-81.76$ & $0.78$ & $-2.25$ & $0.00$ & $15.17$ & $15.14$ \\
693659006674064768 & $140.24$ & $25.53$ & $6.49$ & $1.05$ & $-2.76$ & $0.01$ & $14.95$ & $14.99$ \\
1612085914078414976 & $214.84$ & $58.86$ & $-147.78$ & $1.16$ & $-2.56$ & $0.05$ & $15.74$ & $15.87$ \\
1616913045002580992 & $220.85$ & $58.61$ & $-153.75$ & $1.32$ & $-2.56$ & $0.00$ & $15.83$ & $15.98$ \\
1613056920284943616 & $228.84$ & $57.60$ & $-163.90$ & $1.86$ & $-2.58$ & $0.01$ & $15.79$ & $15.93$ \\
1574576865170810880 & $187.38$ & $56.06$ & $-119.88$ & $0.96$ & $-2.06$ & $0.01$ & $15.52$ & $15.70$ \\
1566732502541484928 & $200.54$ & $58.09$ & $-135.98$ & $1.24$ & $-2.43$ & $0.06$ & $15.58$ & $15.67$ \\
\bottomrule
\end{tabular}

    \caption{DESI-DR2 observations of the BHB stars in GD-1 region as identified in Section \ref{sec:cmd}. The full machine-readable table with additional columns is available on Zenodo (see Data Availability).}
    \label{tab:bhb-members}
\end{table*}

\section{DESI Observations of the GD-1 Stream and Sample Completeness}
\label{sec:appendix_completeness}

Because DESI observations of GD-1 span commissioning, survey validation, and the first three years of the main survey, the spectroscopic coverage of the stream is intrinsically heterogeneous. GD-1 extends over more than $100^\circ$ on the sky, and stars along the stream have been observed in multiple DESI \emph{surveys} (\texttt{SV1}--\texttt{SV3}, \texttt{main}, and \texttt{special}) and \emph{programs} (\texttt{bright}, \texttt{dark}, \texttt{backup}). Each survey--program combination corresponds to a different targeting strategy and set of observing conditions \citep{Koposov_2025}. Although this complexity makes the selection function non-uniform on the sky, our main scientific conclusions do not rely on a uniform or volume-complete sample. In this Appendix we summarize the GD-1 coverage in various DESI survey programs in DR2, and quantify the completeness of our working sample.

\begin{figure*}
    \centering
    \includegraphics[width=\linewidth]{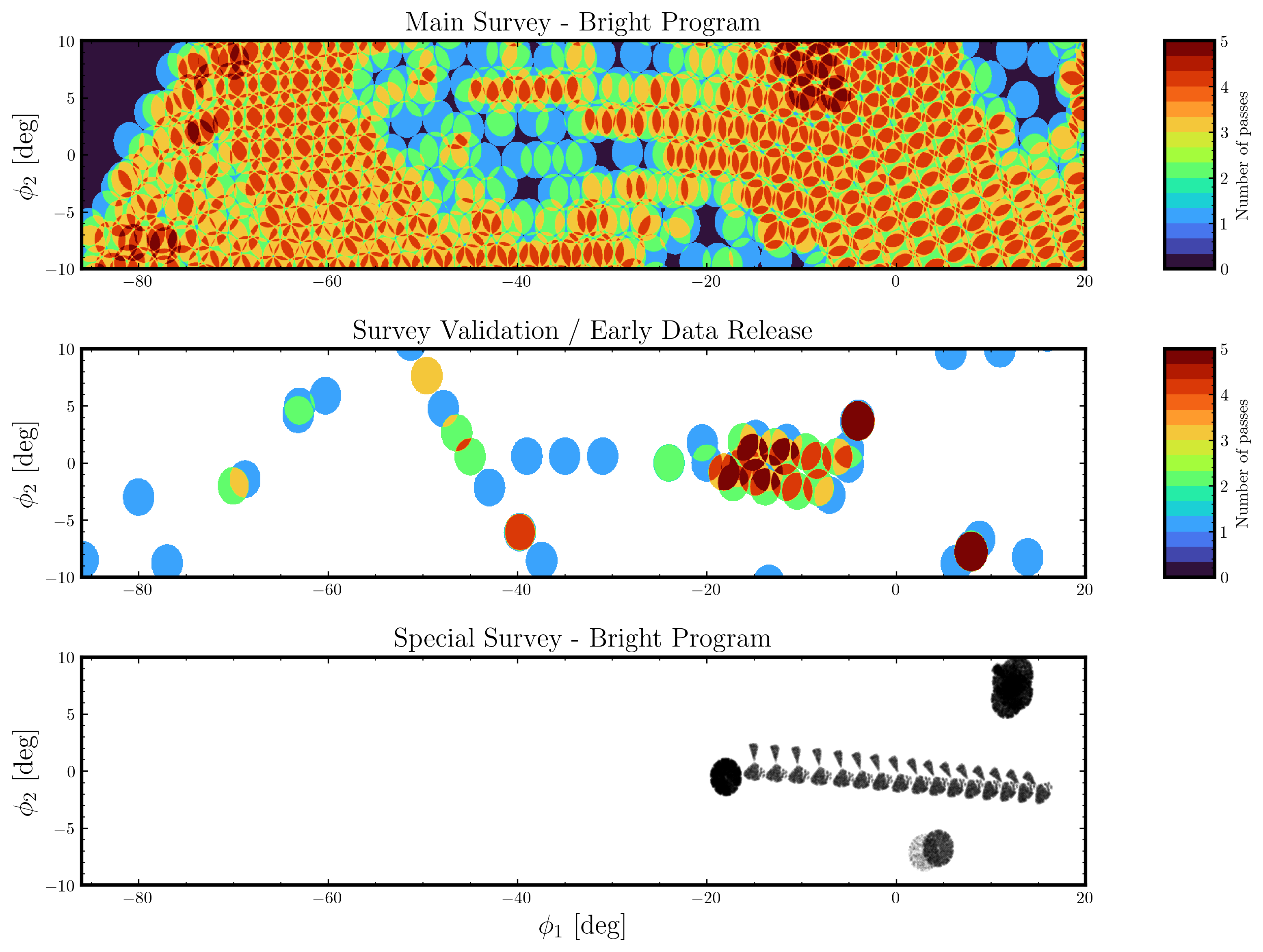}
    
    \caption{
    DESI tiling of the GD-1 region in the stream-aligned coordinate system $(\phi_1,\phi_2)$.
    \emph{Top:} number and locations of passes from the main-survey \texttt{bright} program contributing to the DESI DR2 footprint; colors indicate the total number of bright-time passes at each position along the stream.
    \emph{Middle:} number of DESI tiles from Survey Validation (\texttt{SV1}--\texttt{SV3}) and the Early Data Release used in \citet{valluri_2025}, illustrating the legacy GD-1 coverage inherited from EDR.
    \emph{Bottom:} stars observed in the GD-1 region as part of the \texttt{special} survey under bright conditions, including the deep ($\sim 2$\,hr) DESI-II preparation tile near $\phi_1 \sim -19^\circ$ and the 16-tile GD-1 mini-survey obtained when four of the ten spectrographs were operational.
    Together, these panels highlight the strongly non-uniform but complementary DESI coverage of GD-1 across different surveys and programs in DR2.
    }
    \label{fig:tiles}
\end{figure*}

Figure~\ref{fig:tiles} shows the DESI tiles overlapping the GD-1 stream in the DR2 footprint. Among the 679 stars in the ``DESI complete selection'' sample defined in Section~\ref{sec:mem_sel_summary}, approximately $60\%$ were observed in the \texttt{main bright} program. This is expected: the bright program is the primary mode of the MWS and provides the majority of DESI stellar exposures. The top panel of Figure~\ref{fig:tiles} displays the corresponding tiling pattern. Because DR2 corresponds to the mid-point of the five-year main survey, the nominal five-pass coverage is not yet achieved everywhere, and the bright-time completeness is visibly non-uniform along the stream.

Roughly $29\%$ of the GD-1 sample was observed as a part of the \texttt{special} programs. These include two sets of dedicated GD-1 observations. First, a deep tile with a long $\sim 2$\,hr effective exposure time centered near $\phi_1 \sim -19^\circ$ was observed to test DESI performance at the faint end ($r\gtrsim20$) in preparation for DESI-II. Second, a 16-tile GD-1 program with effective exposure times of $\sim 5$\,min per tile was executed when only 4 of the 10 spectrographs were operational, resulting in the distinctive tiling geometry shown in the bottom panel of Figure~\ref{fig:tiles}. Together, these special tiles significantly increase the depth and local sampling of GD-1 near the central portion of the stream.

The middle panel of Figure~\ref{fig:tiles} highlights the contribution from Survey Validation (\texttt{SV1}--\texttt{SV3}), which contributes $\sim 20\%$ of the final sample. These data include the GD-1 observations analyzed in the DESI Early Data Release \citep{valluri_2025}. Because SV programs span bright, dark, and ``other'' conditions, they help to fill in several regions along GD-1 that are incompletely covered by the main survey alone.

In addition, $\sim 24\%$ and $\sim 16\%$ of the sample were observed in the \texttt{main dark} and \texttt{main backup} programs, respectively. Although the main dark program is primarily designed for extragalactic targets, MWS stars are allocated low-priority fibers and thus a subset of GD-1 stars are observed in dark time. The backup program operates during twilight and poor weather and predominantly targets bright stars ($r < 16$) \citep{Dey_2025}, but still contributes a useful fraction of GD-1 members. The sum of these fractions exceeds $100\%$ because roughly two-thirds of the GD-1 stars in our catalog have multi-epoch DESI spectroscopy across different surveys and programs. A systematic analysis of binaries and radial-velocity variability using this multi-epoch information is beyond the scope of this paper and will be pursued in future work.

\begin{figure*}
    \centering
    \includegraphics[width=\linewidth]{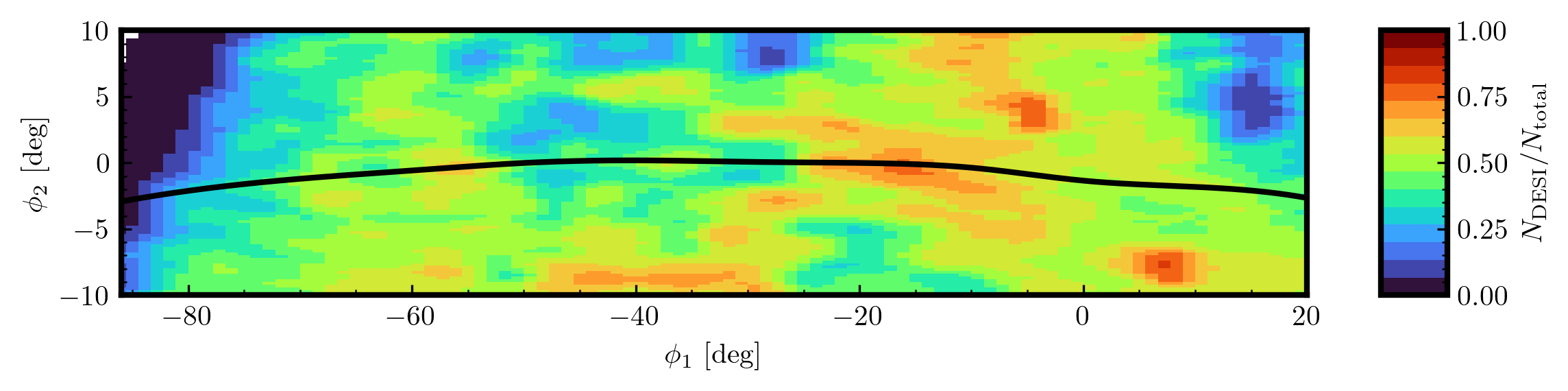}
    \caption{
    Spectroscopic completeness of DESI observations in the GD-1 region.
    The color scale shows, in each spatial bin, the fraction of {\it Gaia}+LS stars (selected with the same color-magnitude and proper-motion criteria and limited to the magnitude range $16 < r < 19$) that have DESI spectroscopy in our complete selection sample.
    Completeness closely follows the tiling pattern of the main-survey \texttt{bright} program: areas with multiple bright-time passes show higher completeness, while gaps in the bright tiling appear as regions of low completeness.
    The particularly high completeness around stream longitudes between roughly $-20$ and $-10$ degrees reflects the combined effect of Survey Validation data and the dedicated GD-1 special-survey fields in this part of the stream as shown in Figure \ref{fig:tiles}.
        }
    \label{fig:completeness}
\end{figure*}

Our goal in this work is to construct the largest, cleanest spectroscopic catalog of GD-1 members available in DESI-DR2. To this end, we merge observations from all relevant surveys and programs and retain a single best measurement per star (i.e., entries with \texttt{PRIMARY = TRUE}). We then quantify the completeness of this merged sample in Figure~\ref{fig:completeness}. Here, the completeness at a given position along the stream is defined as
\begin{equation}
    f_{\rm comp} = \frac{N_{\rm DESI}}{N_{\rm {\it Gaia}+DECaLS}},
\end{equation}
where $N_{\rm DESI}$ is the number of stars in the DESI complete selection sample and $N_{\rm {\it Gaia}+DECaLS}$ is the number of {\it Gaia}+DECaLS stars satisfying the same color--magnitude and proper-motion criteria, both restricted to $16 < r < 19$\footnote{We note, however, that in various programs such as \texttt{main dark} and \texttt{special}, targets fainter than $r=19$ are also observed, extending beyond the main MWS sample.}.

The completeness pattern closely follows the bright-program tiling: regions with multiple bright-time passes exhibit higher completeness, while gaps in the bright tiling map directly into lower spectroscopic sampling. The enhanced completeness around $-20^\circ < \phi_1 < -10^\circ$ is driven by the overlap of SV observations and the dedicated special-survey tiles described above. Once the DESI main survey is completed in 2029, we expect the GD-1 spectroscopic sample in the range $16 < r < 19$ to at least double in size, and extended bright- and dark-time coverage will increase the number of faint GD-1 members with high-quality spectroscopy.

\bibliography{bib}
\bibliographystyle{aasjournalv7}

\end{document}